# Fluid Agency in AI Systems: A Case for Functional Equivalence in Copyright, Patent, and Tort

Anirban Mukherjee

Hannah Hanwen Chang

Anirban Mukherjee (anirban@avyayamholdings.com) is Principal at Avyayam Holdings, Singapore. Hannah H. Chang (hannahchang@smu.edu.sg; corresponding author) is Associate Professor of Marketing at the Lee Kong Chian School of Business, Singapore Management University. This research was supported by the Ministry of Education (MOE), Singapore, under its Academic Research Fund (AcRF) Tier 2 Grant, No. MOE-T2EP40221-0008.

## Abstract

Modern Artificial Intelligence (AI) systems lack human-like consciousness or culpability, yet they exhibit *fluid agency*: behavior that is (i) *stochastic* (probabilistic and path-dependent), (ii) *dynamic* (co-evolving with user interaction), and (iii) *adaptive* (able to reorient across contexts). Fluid agency generates valuable outputs but collapses attribution, irreducibly entangling human and machine inputs. This fundamental *unmappability* fractures doctrines that assume traceable provenance—authorship, inventorship, and liability—yielding ownership gaps and moral "crumple zones."

This Article argues that only *functional equivalence* stabilizes doctrine. Where provenance is indeterminate, legal frameworks must treat human and AI contributions as equivalent for allocating rights and responsibility—not as a claim of moral or economic parity but as a pragmatic default. This principle stabilizes doctrine across domains, offering administrable rules: in copyright, vesting ownership in human orchestrators without parsing inseparable contributions; in patent, tying inventor-of-record status to human orchestration and reduction to practice, even when AI supplies the pivotal insight; and in tort, replacing intractable causation inquiries with enterprise-level and sector-specific strict or no-fault schemes. The contribution is both descriptive and normative: fluid agency explains why origin-based tests fail, while functional equivalence supplies an outcome-focused framework to allocate rights and responsibility when attribution collapses.

# INTRODUCTION

Foundational legal doctrines of authorship, inventorship, and liability rest on a critical, often unstated, assumption: the ability to trace a creative work or a harmful act back to a human agent. For centuries, this principle of attribution has provided a stable basis for allocating rights and responsibilities.

Modern Artificial Intelligence (AI) systems are beginning to fracture this bedrock. Early 2025 saw the emergence of systems that execute complex workflows with what we term *fluid agency*: (i) *stochastic*, (ii) *dynamic*, and (iii) *adaptive* action pathways that blur the line between a human creator and her tool.[1] While these systems share roots with what were termed "intelligent agents," they represent a significant leap in sophistication as their predecessors were constrained to narrowly defined tasks and rigid rules. In contrast, modern systems interpret context, adapt strategies, and proactively orchestrate multi-step processes: capabilities that underpin a shift from reactive and advisory functions to proactive execution—e.g., negotiating pricing with suppliers, rerouting shipments to avoid geopolitical disruptions, and recalibrating production schedules in response to fluctuating

---

[1] In the Artificial Intelligence literature, *agency* is the capacity for autonomous goal-directed action. *See, e.g.*, Stan Franklin & Art Graesser, *Is It an Agent, or Just a Program?: A Taxonomy for Autonomous Agents*, INTELLIGENT AGENTS III: AGENT THEORIES, ARCHITECTURE, AND LANGUAGES*,* 1996, at 21, 25 (defining an autonomous agent as "a system situated within and a part of an environment that senses that environment and acts on it, over time, in pursuit of its own agenda and so as to effect what it senses in the future."); Michael Wooldridge & Nicholas R. Jennings, *Intelligent Agents: Theory and Practice*, 10 KNOWL. ENG'G REV. 115, 116 (1995) (identifying core properties of agents including *autonomy*—operating "without the direct intervention of humans"—and *pro-activeness*, i.e., goal-directed behavior "by taking the initiative"); J. M. Beer, A. D. Fisk & W. A. Rogers, *Toward a Framework for Levels of Robot Autonomy in Human-Robot Interaction*, 3 J. HUM.-ROBOT INTERACT. 74, 77 (2014) (defining autonomy in agents as "the extent to which a system can carry out its own processes and operations without external control"); Jeffrey M. Bradshaw, Robert R. Hoffman, Matthew Johnson, & David D. Woods;., *The Seven Deadly Myths of "Autonomous Systems"*, 28 IEEE INTELL. SYST. 54, 56, 60–61 (2013) (distinguishing "self-sufficiency—the capability of an entity to take care of itself" from "self-directedness, or freedom from outside control" and arguing the term "autonomous system" is often misleading given human-machine interdependence).

demand—and are thus foundational to AI's potential.[2] However, they also obscure the boundary between the initiative and actions of an AI and its human user, creating a fundamental attribution challenge.

Consider Deep Research Agents (DRAs)—systems such as OpenAI's Deep Research—which "autonomously orchestrate multi-step web exploration, targeted retrieval, and higher-order synthesis," distilling vast information into analyst-grade reports.[3] DRAs are more than amanuenses[4] but less than collaborators: they exercise agency over the *means* of research but not the *ends*. They independently decide *how* to synthesize information—assessing source credibility, weighing conflicting information, and structuring final reports—tasks previously reserved for human judgment.[5] Yet the *why* and the *what for*: the purpose of the research and how it is used remain with their human users. When a DRA's work is folded into a final report, authorship becomes fundamentally unclear: the report reflects human inputs but also the AI's independent choices about sources, weighting, and structure.

---

[2] *See* Yonadav Shavit, Sandhi Agarwal, Miles Brundage, Steven Adler, Cullen O'Keefe, Rosie Campbell, Teddy Lee, Pamela Mishkin, Tyna Eloundou, Alan Hickey, Katarina Slama, Lama Ahmad, Paul McMillan, Alex Beutel, Alexandre Passos, & David G Robinson, *Practices for Governing Agentic AI Systems,* OPENAI*,* (Dec. 14, 2023), https://cdn.openai.com/papers/practices-for-governing-agentic-ai-systems.pdf [https://perma.cc/KKZ6-9TMK](offering a working definition of agentic AI—systems that pursue goals over extended periods without behavior being fully specified in advance—and governance practices including approval gating, action constraints, legibility, and interruptibility); Rina Diane Caballar, *What Are AI Agents?*, IEEE SPECTRUM (Nov. 19, 2024), https://spectrum.ieee.org/ai-agents [https://perma.cc/NH4T-S9DW] (plain-language overview showing how AI agents decompose goals into subtasks and act via tools—"humans set the goal, and agents figure out ... the best course of action").

[3] *See* Mingxuan Du, Benfeng Xu, Chiwei Zhu, Xiaorui Wang & Zhendong Mao, *DeepResearch Bench: A Comprehensive Benchmark for Deep Research Agents*, ARXIV (June 13, 2025), https://arxiv.org/pdf/2506.11763 [https://perma.cc/486T-NNUH] (defining the capabilities of DRAs and introducing a 100-task PhD-level benchmark to evaluate them); *see also* OPENAI, *Introducing Deep Research* (Feb. 2, 2025), https://openai.com/index/introducing-deep-research/ [https://perma.cc/XQ5J-5VJN] (announcing OpenAI's *Deep Research* as a DRA designed to produce reports "like a research analyst").

[4] *See* Jane C. Ginsburg & Luke Ali Budiardjo, *Authors and Machines*, 34 BERK. TECH. L.J. 343, 344 (2019).

[5] *See* Deepak Bhaskar Acharya, Karthigeyan Kuppan & B. Divya, *Agentic AI: Autonomous Intelligence for Complex Goals—A Comprehensive Survey*, 13 IEEE ACCESS 18912, 18913 (2025) (describing autonomous systems capable of accomplishing complex, long-term tasks without direct human supervision).

This ambiguity arises because a DRA's agency is partial—distinct from human agency, traditional tools with no agency, and hypothetical, fully autonomous artificial general intelligence (AGI) with full agency. Unlike humans, it lacks the consciousness and culpability inherent in human agency,[6] which forms the basis of legal accountability and the attribution of rights.[7] Unlike simple tools, it is emulative of human agency—its behavior derives from processes optimized for reward signals aligned with human decision-making patterns. And unlike both conscious beings and fully autonomous AGI that pursue their *own* agenda,[8] it operates under the goals and constraints set by a human user.[9]

Fluid agency introduces novel challenges for existing legal and policy frameworks. The issue stems from the fundamental *unmappability* of roles and contributions in intertwined human-AI processes.[10] For instance, as the DRA example illustrates, contributions in a creative collaboration can defy categorization as originating from human or AI sources.[11] This renders traditional legal frameworks that presuppose a divisible chain of creation impracticable.[12]

---

[6] Moral agency, as understood in Western law and philosophy, requires a conscious self with subjective experience, intent, and the capacity for rational judgment. *See, e.g.*, Markus Schlosser, *Agency*, *in* STAN. ENCYC. PHIL. (Edward N. Zalta ed. October 2019),.

[7] *See, e.g.*, Cruzan v. Dir., Mo. Dep't of Health, 497 U.S. 261, 279 (1990) (recognizing—and assuming for purposes of the case—that a competent person has a constitutionally protected liberty interest in refusing unwanted medical treatment); RESTATEMENT (SECOND) OF CONTRACTS § 15(1)(a) (AM. L. INST. 1981) (defining incapacity in terms of whether a person is "unable to understand in a reasonable manner the nature and consequences of the transaction").

[8] *See, e.g.* Franklin & Graesser, *supra* note 1, at 25; Bartosz Brożek & Marek Jakubiec, *On the Legal Responsibility of Autonomous Machines*, 25 ARTIF. INTELL. & L. 293, 294 (2017).

[9] Parallels can be drawn to both non-rational "actors" (e.g., domesticated animals) whose acts are imputed to the owner and to corporations whose "agency" does not presuppose will. *See, e.g.*, 2 WILLIAM BLACKSTONE, COMMENTARIES ON THE LAWS OF ENGLAND, ch. XXV (1765–69) (classifying animals as chattels); 3 *id.* bk. III, ch. XII (discussing liability for animal trespass); RESTATEMENT (SECOND) OF TORTS § 509 (AM. L. INST. 1977) (imposing strict liability on a possessor of a domestic animal that they know or have reason to know has abnormally dangerous propensities). But unlike non-rational actors, AI systems demonstrate sophisticated rationalizing, and unlike corporations, they are not governed by rational actors.

This Article contends that only the principle of *functional equivalence* stabilizes authorship, inventorship, and liability when fluid agency renders origins unmappable. How can doctrine distinguish between "human" and "machine" when, by nature, contributions and roles are irreducibly entangled? Any rule that presupposes separable origins either (i) fails outright when the boundary cannot be drawn, or (ii) reintroduces a threshold test for separability that is unadministrable in typical workflows. Functional equivalence[13] is the minimal, domain-general principle that dispenses with origin tracing. We therefore propose that where provenance is indeterminate after reasonable provenance practices, legal frameworks should treat human and AI contributions as equivalent for the purpose of allocating rights and responsibility as a pragmatic default, and not a claim of moral or ontological parity.

---

[10] By unmappability, we mean the practical impossibility of reliably attributing specific elements of an output to either human or machine.

[11] Some scholars argue against framing human–machine interactions as collaborative, asserting that true collaboration requires shared intentionality, moral agency, and the ability to co-determine objectives—attributes they deem absent in AI, which they view as heteronomous tools (i.e., governed externally rather than by self-determination). *See, e.g.*, K. D. Evans, S. A. Robbins & J. J. Bryson, *Do We Collaborate with What We Design?*, 17 TOP. COGN. SCI. 392 (2025). However, this critique presumes that machines lack the capacity to participate in open-ended creative processes. Modern AI subverts this premise. For example, when tasked with producing a climate report, the AI might autonomously refocus the analysis from mitigation costs to adaptation ethics based on its assessment of emerging scholarship. While the human sets the broad mandate, the AI dynamically determines the specific objectives and methodological trajectory—a form of *procedural co-determination* that blurs the intentional hierarchy (namely, that humans have intentions while a mere tool does not) central to heteronomy critiques. This fluid renegotiation of sub-goals defies clean attribution, making "collaboration" less a metaphor and more a functional descriptor of such creative entanglement.

[12] Authors such as Annemarie Bridy consider the case where "digital works (i.e., software programs) will, relatively autonomously, produce other works that are indistinguishable from works of human authorship." *See* Annemarie Bridy, *Coding Creativity: Copyright and the Artificially Intelligent Author*, 5 STAN. TECH. L. REV. 1, ¶ 7 (2012). However, they maintain the assumption that human and machine contributions can be separated. This holds, for example, when the output in question is developed using an AI whose actions are predictable (e.g., a text-to-image AI will always generate a different output—a distinct image—but will always undertake the same action—it will generate an image). Modern AI, in contrast, brings to the fore scenarios where, in addition to an AI's outputs being indistinguishable from human outputs, its pathway to those outputs is unpredictable and the drivers of those outputs become irreducibly entangled between a human and her AI.

This Article proceeds as follows. *Part I* establishes the concept of fluid agency. *Part II* describes the resulting unmappability of contributions and roles, which destabilizes current legal and policy paradigms. *Part III* examines implications for copyright law; *Part IV* addresses inventorship challenges in patent law; and *Part V* considers liability allocation in tort law. *Part VI* introduces and defends functional equivalence. *Part VII* concludes by outlining practical implementation pathways, discussing the principle's limitations, and exploring its broader societal implications.

# I. FLUID AGENCY

Prevailing discourse on the implications of AI for copyright, patent, and tort law has coalesced around a binary conceptualization of agency, framing AI systems as either rigidly predictable tools or sovereign actors. Modern AI disrupts this dichotomy by introducing an intermediate form of agency wherein, like tools, AIs work towards human-set goals but, like sovereign actors, exercise significant independence. In this Part, we characterize this agency and trace its implications for copyright, patent, and tort law.

---

[13] "Functional equivalence" has intellectual precedents across disciplines. In linguistics and sociology, it denotes how different forms or structures can fulfill the same essential function. *See, e.g.*, EUGENE A. NIDA, TOWARD A SCIENCE OF TRANSLATING 159 (1964) (distinguishing between formal and dynamic/functional equivalence in translation); ROBERT K. MERTON, SOCIAL THEORY AND SOCIAL STRUCTURE 90–91, 106 (1968 ed.) (discussing functional alternatives in social institutions). In law, a parallel principle is well-established in electronic-commerce frameworks, which grant electronic records and signatures the same legal validity as their paper-based counterparts. *See, e.g.*, G.A. Res. 60/21 art. 9, United Nations Convention on the Use of Electronic Communications in International Contracts (Dec. 9, 2005); UNIF. ELEC. TRANSACTIONS ACT § 7 (UNIF. L. COMM'N 1999) (adopted by 49 states, the District of Columbia, Puerto Rico, and the U.S. Virgin Islands; New York has enacted its own statute, the Electronic Signatures and Records Act (ESRA)). However, this concept has not been systematically applied to resolve the attribution crises in *authorship, inventorship, or liability*—the doctrinal gaps exposed by *fluid agency*. We adapt it here as a pragmatic legal principle: for the purpose of assigning rights and liability, human and AI contributions should be treated as interchangeable, not due to any moral or ontological parity, but as a necessary response to the practical impossibility of disentangling their origins.

Automation—rule-bound systems designed for deterministic, interpretable execution of predefined tasks, such as TurboTax's form-filling algorithms or will preparation software that mechanizes rote legal drafting—exemplifies the first perspective of intelligent systems as tools.[14] Such systems presuppose and require human oversight at every critical juncture, ensuring legitimacy through "humans in the loop" who intervene to resolve ambiguities or edge cases.[15]

Traditional generative AI, where a user maintains almost complete control over an AI's actions through iterative prompting and output curation, fits this mode.[16] For instance, a text-to-image system like DALL-E generates outputs in response to a human-provided prompt. Although the output image is unpredictable, a function of the prompt and random noise, the AI's action is perfectly predictable (it generates images in response to prompts).[17] A human can then experiment with prompts, reviewing and selecting among the generated images. The role of each—the human as a provider of prompts and a curator of images, and the AI as an image generator—remains well defined.[18]

---

[14] Frank Pasquale, *A Rule of Persons, Not Machines: The Limits of Legal Automation*, 87 GEO. WASH. L. REV. 1, 15–20 (2019) (arguing that legal automation thrives only when confined to "rules of recognition" with clear, human-verifiable inputs and outputs, but falters in interpretive domains requiring contextual judgment); Richard M. Re & Alicia Solow-Niederman, *Developing Artificially Intelligent Justice*, 22 STAN. TECH. L. REV. 242, 254–56, 265–66 (2019) (analyzing how AI adjudication favors "codified justice" over "equitable justice" and raising due-process concerns regarding incomprehensibility, rather than simply advocating for interpretability).

[15] Rebecca Crootof, Margot E. Kaminski & W. Nicholson Price II, *Humans in the Loop*, 76 VAND. L. REV. 429, 450–60 (2023) (analyzing how law requires humans "in, on, or off the loop" and cautioning that merely adding a human does not guarantee accountability); Ryan Calo & Danielle Keats Citron, *The Automated Administrative State: A Crisis of Legitimacy*, 70 EMORY L.J. 797, 820–30 (2021) (arguing that opacity and a lack of reviewability in automated systems undermine administrative legitimacy and advocating for explainability, contestability, and auditability).

[16] *See* Pamela Samuelson, *Generative AI Meets Copyright*, 381 SCIENCE 158, 158–61 (2023) (describing how users of traditional generative AI direct outputs through prompting and curation).

[17] *See* Aditya Ramesh, Prafulla Dhariwal, Alex Nichol, Casey Chu & Mark Chen, *Hierarchical Text-Conditional Image Generation with CLIP Latents*, ARXIV (Apr. 13, 2022), https://arxiv.org/pdf/2204.06125 [https://perma.cc/T6HQ-LGP9] (describing the DALL-E 2

Because attribution is clear, liability traces to human operative decisions, authorship to human creative direction, and inventorship to human conception. Much as a calculator bears no blame for a faulty equation, a word processor earns no byline for a manuscript, and a computer-aided design program claims no credit for a blueprint, the machine is removed from the equation.

By contrast, complete, sovereign autonomy, where AI pursues independent goals without human input (i.e., AGI) renders attribution moot as machines become de facto independent moral agents.[19] Scholarship on AGI debates radical reforms, such as *sui generis* rights for machine "persons" or no-fault liability regimes to bridge the "responsibility gap" that arises from fully autonomous acts.[20] It proposes "artificially intelligent justice," shifting from discretionary "equitable justice" to predictable "codified justice,"[21] with "robot judges" and hybrid human-AI courts.[22]

---

architecture where the generative process is fixed, but outputs vary based on stochastic sampling of the latent space).

[18] *Cf.* Ginsburg & Budiardjo, *supra* note 4, at 441–42 (arguing that in prompt-based generative systems, human curation of outputs can establish authorship precisely because the machine's role is mechanistically confined, preserving a clear attribution pathway).

[19] *See* NICK BOSTROM, SUPERINTELLIGENCE: PATHS, DANGERS, STRATEGIES 22–25 (2014) (envisioning AGI as self-improving systems capable of recursive goal formulation, raising existential risks from unaligned agency).

[20] *See* Andreas Matthias, *The Responsibility Gap: Ascribing Responsibility for the Actions of Learning Automata*, 6 ETHICS & INFO. TECH. 175, 178–82 (2004) (coining the "responsibility gap"); *see also* Resolution of 16 February 2017 with recommendations to the Commission on Civil Law Rules on Robotics, EUR. PARL. DOC. P8_TA(2017)0051, 2015/2103(INL), ¶ 59(f) (suggesting the Commission consider creating "a specific legal status for robots . . . as having the status of electronic persons").

[21] Re & Solow-Niederman, *supra* note 14, at 252–55.

[22] *See, e.g.*, Eugene Volokh, *Chief Justice Robots*, 68 DUKE L.J. 1135, 1144–50 (2019); Benjamin Minhao Chen, Alexander Stremitzer & Kevin Tobia, *Having Your Day in Robot Court*, 36 HARV. J.L. & TECH. 127, 132–36 (2022).

These framings treat AI as either a highly advanced tool for human-directed tasks, almost entirely subservient to human will, or a complete replacement for the human, rather than as a limited partner in a dynamic, co-evolving process. Yet, modern AI is frequently neither. Even as it exercises considerable independence in means, its overarching objectives and constraints remain set by a human user (e.g., "write a research paper, citing only peer-reviewed papers" or "optimize supply chain costs without introducing new vendors"). Its agency is fluid as its action processes are: (i) stochastic, as its multi-step pathways are probabilistic; (ii) dynamic, as it responds to inputs within the context of prior interactions; and (iii) adaptive, as it learns implicitly from user feedback, internalizing preferences without explicit instruction.

Its stochasticity arises from its foundation in generative models. Unlike symbolic AI, which follows explicit, deterministic rules, modern AI's operational pathways are assembled probabilistically.[23] Randomness manifests at three levels: (1) action selection at each decision node, (2) adaptation to workflow states and prior interactions, and (3) interpretive variance in processing user feedback. Outcomes are the result of a multi-step process where at each step, an AI can branch along divergent paths. The choice probabilities compound to produce *chaotic divergence*.[24] In a computational analog of the

---

[23] For foundational work on the stochasticity that underlies generative models, *see, e.g.*, Diederik P. Kingma & Max Welling, *An Introduction to Variational Autoencoders*, 12 FOUND. TRENDS MACH. LEARN. 307 (2019); Ian J. Goodfellow, Jean Pouget-Abadie, Mehdi Mirza, Bing Xu, David Warde-Farley, Sherjil Ozair, Aaron Courville & Yoshua Bengio, *Generative Adversarial Nets*, *in* 27 ADVANCES IN NEURAL INFO. PROCESSING SYS. 2672, 2672–80 (2014); Jonathan Ho, Ajay Jain & Pieter Abbeel, *Denoising Diffusion Probabilistic Models*, *in* 33 ADVANCES IN NEURAL INFO. PROCESSING SYS. 6840, 6840–51 (2020); Tom B. Brown, Benjamin Mann, Nick Ryder, Melanie Subbiah, Jared Kaplan, Prafulla Dhariwal, Arvind Neelakantan, Pranav Shyam, Girish Sastry, Amanda Askell, Sandhini Agarwal, Ariel Herbert-Voss, Gretchen Krueger, Tom Henighan, Rewon Child, Aditya Ramesh, Daniel M. Ziegler, Jeffrey Wu, Clemens Winter, Christopher Hesse, Mark Chen, Eric Sigler, Mateusz Litwin, Scott Gray, Benjamin Chess, Jack Clark, Christopher Berner, Sam McCandlish, Alec Radford, Ilya Sutskever & Dario Amodei, *Language Models Are Few-Shot Learners*, ARXIV (May 28, 2020), https://arxiv.org/pdf/2005.14165 [https://perma.cc/BXX6-MW5G]; Yoshua Bengio & Yann LeCun, *Scaling Learning Algorithms toward AI*, *in* LARGE-SCALE KERNEL MACHINES 321, 321–360 (Léon Bottou et al. eds., 2007).

[24] Mechanisms such as epsilon-greedy exploration, where an agent randomly selects a non-optimal action with a given probability to explore alternative strategies, amplify this stochastic nature. *See, e.g.*, RICHARD S. SUTTON & ANDREW G. BARTO, REINFORCEMENT LEARNING: AN INTRODUCTION, 31-52 (2d ed. 2018).

"butterfly effect," microscopic differences in initial conditions lead to macroscopic divergence.[25]

An AI responds to current user inputs within the context of the user's prior inputs and the AI's own prior outputs, creating a dynamic feedback loop. Its operational boundaries are molded by the ongoing dialogue. For instance, negative user feedback may lead an AI to curtail its autonomy, while positive feedback may encourage greater initiative.[26] This dynamic quality distinguishes modern AI from static tools (e.g., text-to-image AI systems such as DALL-E) where each input triggers a relatively isolated response.

Finally, modern AI's fluid agency is adaptive. When considering machine creativity, Ginsburg and Budiardjo write, "[t]he computer scientist who succeeds at the task of 'reduc[ing] [creativity] to logic' does not generate new 'machine' creativity—she instead builds a set of instructions to codify and simulate 'substantive aspect[s] of human [creative] genius,' and then commands a computer to faithfully follow those instructions."[27] Ginsburg and Budiardjo assume an AI must be programmed to be creative. While this was true for traditional symbolic AI, modern AI is now almost exclusively programmed to learn from data and user interactions rather than with a specific body of knowledge. Its programmers do not code explicit instructions; rather, the AI develops its

---

[25] *Cf.* Xuezhi Wang, Jason Wei, Dale Schuurmans, Quoc Le, Ed Chi, Sharan Narang, Aakanksha Chowdhery & Denny Zhou, *Self-Consistency Improves Chain of Thought Reasoning in Language Models*, ARXIV (Mar. 21, 2022), https://arxiv.org/pdf/2203.11171 [https://perma.cc/R7SX-TRTJ] (demonstrating that sampling diverse reasoning paths from a language model produces a range of different outcomes); *see also* Edward N. Lorenz, *Deterministic Nonperiodic Flow*, 20 J. ATMOSPHERIC SCI. 130, 141 (1963).

[26] *See* Zhehao Zhang, Ryan A. Rossi, Branislav Kveton, Yijia Shao, Diyi Yang, Hamed Zamani, Franck Dernoncourt, Joe Barrow, Tong Yu, Sungchul Kim, Ruiyi Zhang, Jiuxiang Gu, Tyler Derr, Hongjie Chen, Junda Wu, Xiang Chen, Zichao Wang, Subrata Mitra, Nedim Lipka, Nesreen K. Ahmed & Yu Wang, *Personalization of Large Language Models: A Survey*, TRANS. MACH. LEARN. RES. (2025), https://openreview.net/pdf?id=tf6A9EYMo6 [https://perma.cc/67EN-PRL6] (describing, in §§ 3.1–3.2, how LLMs form a "relationship memory" through dynamic feedback loops, evolving their understanding of a task via in-context learning and implicit preference alignment).

[27] Ginsburg & Budiardjo, *supra* note 4, at 350–51.

faculties implicitly through mechanisms like reinforcement learning from human feedback.[28]

This training makes its behavior contextual. The level of human control is never fixed as AI's planning, execution, and outputs vary significantly across interactions and tasks, adapting to the preferences and knowledge structures of its user. Such AI can exhibit emergent behavior—complex, unpredictable patterns that result from its training, inference, and model structure, and relate to its adaptation during use.[29]

This capacity for adaptive behavior makes AI intrinsically unpredictable, complicating direct human control. Moreover, adaptation can be a two-way street whereby an AI internalizes a user's preferences through iterative interaction, while the user adapts their thinking and creative processes to the AI's outputs.[30] Such a co-evolutionary recursively adaptive feedback loop further amplifies the causal entanglement that—as the next Part shows—makes attribution fundamentally unmappable.

---

[28] *See, e.g.*, Yuntao Bai, Saurav Kadavath, Sandipan Kundu, Amanda Askell, Jackson Kernion, Andy Jones, Anna Chen, Anna Goldie, Azalia Mirhoseini, Cameron McKinnon, Carol Chen, Catherine Olsson, Christopher Olah, Danny Hernandez, Dawn Drain, Deep Ganguli, Dustin Li, Eli Tran-Johnson, Ethan Perez, Jamie Kerr, Jared Mueller, Jeffrey Ladish, Joshua Landau, Kamal Ndousse, Kamile Lukosuite, Liane Lovitt, Michael Sellitto, Nelson Elhage, Nicholas Schiefer, Noemi Mercado, Nova DasSarma, Robert Lasenby, Robin Larson, Sam Ringer, Scott Johnston, Shauna Kravec, Sheer El Showk, Stanislav Fort, Tamera Lanham, Timothy Telleen-Lawton, Tom Conerly, Tom Henighan, Tristan Hume, Samuel R. Bowman, Zac Hatfield-Dodds, Ben Mann, Dario Amodei, Nicholas Joseph, Sam McCandlish, Tom Brown & Jared Kaplan, *Constitutional AI: Harmlessness from AI Feedback*, ARXIV (Dec. 15, 2022), https://arxiv.org/pdf/2212.08073 [https://perma.cc/73K4-3VF3] (explaining how reinforcement mechanisms allow an AI to adapt its behavior based on feedback against a set of principles, leading to emergent behaviors that evolve without direct programming).

[29] *See, e.g.*, MELANIE MITCHELL, COMPLEXITY: A GUIDED TOUR 3–14 (2009). While symbolic AI can exhibit some unexpected behaviors, the scale and nature of emergent behavior in modern AI, driven by its learning mechanisms, are qualitatively different.

[30] *See* Martin J. Pickering & Simon Garrod, *Toward a Mechanistic Psychology of Dialogue*, 27 BEHAV. & BRAIN SCI. 169 (2004); Florent Vinchon, Todd Lubart, Sabrina Bartolotta, Valentin Gironnay, Marion Botella, Samira Bourgeois-Bougrine, Jean-Marie Burkhardt, Nathalie Bonnardel, Giovanni Emanuele Corazza, Vlad Glăveanu, Michael Hanchett Hanson, Zorana Ivcevic, Maciej Karwowski, James C. Kaufman, Takeshi Okada, Roni Reiter-Palmon & Andrea Gaggioli, *Artificial Intelligence & Creativity: A Manifesto for Collaboration*, 57 J. CREATIVE BEHAV. 472 (2023).

## II. UNMAPPABILITY

At the ends of the agency spectrum, attributing rights and responsibilities is doctrinally straightforward. With negligible agency, the AI is a mere tool; control, rights, and responsibility for its outputs and actions rest entirely with the human user. With sovereign agency, the AI is the sole actor, bearing full control and, potentially, complete legal accountability. Fluid agency, however, operates in the space in between. Outputs emerge from a recursive co-evolution[31] that irreducibly entangles human and machine inputs, making it practically impossible to trace a specific outcome—whether a creative work or a harmful act—back to a specific origin.

Suppose a user tasks a DRA (such as Deep Research) with a strategic research objective. The user defines the topic, while the DRA manages the execution, identifying relevant publications and synthesizing findings into a structured report. The DRA autonomously determines the appropriate analytical frameworks—a stochastic choice influenced by its initial retrievals. Dynamically, it refines its approach based on user feedback. For example, if a user consistently prioritizes peer-reviewed articles, the DRA *adaptively* learns to favor such sources, even without direct instruction, effectively internalizing the user's preferences.[32] This learning also extends further: a scholar who previously emphasized comparative constitutional law in their prompts may find the DRA autonomously expanding its analysis to include foreign jurisprudence—not because the scholar explicitly requested it, but because the system is extrapolating from the user's demonstrated preferences.

---

[31] The concept of a recursive co-evolution between human and AI finds conceptual analogs in several social science theories. Actor-network theory (ANT), which rejects hierarchical distinctions between human and non-human "actants," provides a particularly apt framework. *See* BRUNO LATOUR, REASSEMBLING THE SOCIAL: AN INTRODUCTION TO ACTOR-NETWORK-THEORY (2005). ANT's symmetrical treatment of agency aligns with the paper's argument that human-AI creative entanglement defies traditional attribution. Similarly, Giddens's structuration theory—where social structures and individual agency recursively shape one another—offers parallels to the fluid agency dynamics described here. *See* ANTHONY GIDDENS, THE CONSTITUTION OF SOCIETY: OUTLINE OF THE THEORY OF STRUCTURATION 1–28 (1984).

[32] Such adaptivity can arise through several mechanisms. *In-context learning* enables the system to draw upon prior interactions—such as user prompts and the model's own outputs. Implicit preference learning, often implemented through reinforcement learning techniques, allows the model to adjust its behavior based on patterns of user approval or correction over time. Explicit adaptation occurs either through direct user instruction or via fine-tuning. During *retrieval-augmented generation*, the system dynamically prioritizes certain information sources. For a comprehensive survey of related approaches, *see*

Given this blending of human and machine initiatives—a direct result of the AI's stochastic choices, dynamic responsiveness, and adaptive learning—is the resulting work a product of human intent, machine autonomy, or an inseparable fusion of both? Contribution and control are blurred across multiple dimensions: *temporally*, the user's oversight dominates during goal-setting while the DRA assumes control during execution; *functionally*, the user defines strategic objectives that the DRA operationalizes through its own tactical decisions; and *interactively*, the entire workflow is subject to continuous, adaptive modification.[33] The demarcation of control and contribution is effectively erased: an insight that arose from an AI's initiative in an interactive research process can perhaps equally be viewed as emergent from the AI mimicking its human user's inputs and feedback as it can from the human user issuing instructions that relate to the same topic based on the AI's prior outputs.

We term this inherent entanglement and the resulting difficulty in definitively tracing origins the *fundamental unmappability*.[34] The creative process that engenders this

---

Zhang et al., *supra* note 26. For a specific example of feedback mechanisms, *see, e.g.*, Bai et al., *supra* note 28.

[33] *See* Zhiheng Xi, Wenxiang Chen, Xin Guo, Wei He, Yiwen Ding, Boyang Hong, Ming Zhang, Junzhe Wang, Senjie Jin, Enyu Zhou, Rui Zheng, Xiaoran Fan, Xiao Wang, Limao Xiong, Yuhao Zhou, Weiran Wang, Changhao Jiang, Yicheng Zou, Xiangyang Liu, Zhangyue Yin, Shihan Dou, Rongxiang Weng, Wensen Cheng, Qi Zhang, Wenjuan Qin, Yongyan Zheng, Xipeng Qiu, Xuanjing Huang & Tao Gui, *The Rise and Potential of Large Language Model Based Agents: A Survey*, ARXIV (Sept. 19, 2023), https://arxiv.org/pdf/2309.07864 [https://perma.cc/S555-FG3A] (describing agentic workflows where the AI autonomously decomposes high-level human goals into sub-tasks and executes them); *see also* Mina Lee, Percy Liang & Qian Yang, *CoAuthor: Designing a Human-AI Collaborative Writing Dataset for Exploring Language Model Capabilities*, ARXIV (Jan. 25, 2022), https://arxiv.org/pdf/2201.06796 [https://perma.cc/K9VK-W7QP] (finding that AI suggestions significantly influence the content and direction of human writing, making it difficult to disentangle contributions).

[34] One might argue that technical solutions such as provenance-tracking or model-logging could resolve this ambiguity. However, a growing body of literature demonstrates that such tools are fundamentally limited in the face of recursive, adaptive systems. They can record a sequence of inputs and outputs, but they cannot map the internal, transformative

unmappability stands in stark contrast to the use of a traditional generative AI by a human. While such systems may produce different outputs (e.g., generate different images from the same prompt), their core action remains predictable (e.g., DALL-E will always generate an image or images in response to a prompt), and they lack the recursively probabilistic internal processes that lead to chaotic divergence. They do not build a rich, evolving contextual understanding based on a long history of user interaction. Nor do their operational boundaries co-evolve in the same ways as their adaptivity is constrained post-training, lacking the capacity for reorientation based on user feedback.[35] Consequently, while these tools present their own set of challenges, they do not engender the same degree of unmappability in contributions and control.

For instance, in the case of *Zarya of the Dawn*, questions regarding authorship were relatively clear because it was undisputed that the author (Kristina Kashtanova) wrote the

---

process by which an AI assimilates human feedback. This process renders outputs as statistically emergent creations—akin to "stochastic parrots" that blend training data without retaining traceable provenance—rather than as linear derivations of specific inputs. *See, e.g.*, Emily M. Bender, Timnit Gebru, Angelina McMillan-Major & Margaret Mitchell, *On the Dangers of Stochastic Parrots: Can Language Models Be Too Big?*, *in* PROC. OF THE 2021 ACM CONF. ON FAIRNESS, ACCOUNTABILITY, & TRANSPARENCY 610, 617 (2021). Moreover, in iterative co-creation, each revision by either human or AI can overwrite or obscure prior contributions, making a clean "chain of custody" practically impossible to maintain across complex workflows. *See, e.g.*, Vinchon et al., *supra* note 30, at 476. Even if a perfect log existed, it could not capture the conceptual origin of an idea, distinguishing the user's guiding intent from the AI's generative execution. The "provenance" thus becomes a history of co-evolutionary entanglement, not a ledger of separable inputs. *See* Iyad Rahwan, Manuel Cebrian, Nick Obradovich, Josh Bongard, Jean-François Bonnefon, Cynthia Breazeal, Jacob W. Crandall, Nicholas A. Christakis, Iain D. Couzin, Matthew O. Jackson, Nicholas R. Jennings, Ece Kamar, Isabel M. Kloumann, Hugo Larochelle, David Lazer, Richard McElreath, Alan Mislove, David C. Parkes, Alex 'Sandy' Pentland, Margaret E. Roberts, Azim Shariff, Joshua B. Tenenbaum & Michael Wellman, *Machine behaviour*, 568 NATURE 477, 483 (2019) (discussing the need to examine "feedback loops between human influence on machine behaviour and machine influence on human behaviour simultaneously" in complex hybrid systems).

[35] *See* Aditya Ramesh, Mikhail Pavlov, Gabriel Goh, Scott Gray, Chelsea Voss, Alec Radford, Mark Chen & Ilya Sutskever, *Zero-Shot Text-to-Image Generation*, 139 PROC. INT'L CONF. MACH. LEARNING 8821 (2021) (explaining that the model's parameters remain frozen post-training, limiting it to a fixed generation pipeline rather than an evolving, adaptive workflow); *cf.* Brown et al., *supra* note 23, at 1880.

text of the graphic novel and used the (traditional) AI image generator Midjourney to create the illustrations.[36] She argued that her detailed text prompts and iterative curation of Midjourney outputs constituted sufficient human authorship to support copyright in the images themselves.[37] That argument was only partially successful.[38] Relying on its human-authorship requirement, the U.S. Copyright Office canceled the original registration and issued a new, narrower registration limited to the human-authored components of the work. The revised certificate protects the text and Kashtanova's selection, coordination, and arrangement of the text and images, but expressly excludes the Midjourney-generated images themselves.[39]

Suppose that instead Kashtanova had discussed the images with an AI with fluid agency.[40] Would the Office have granted her the same copyrights? How much input would she have had to provide the AI, and how substantive would that input need to be, before the Office determined that the "human authorship" requirement was met and that

---

[36] *See* Letter from Robert J. Kasunic, Assoc. Register of Copyrights & Dir. of Registration Pol'y & Prac., U.S. Copyright Office, to Van Lindberg, re: *Zarya of the Dawn* 1–3 (Feb. 21, 2023) [hereinafter *Kasunic Letter*] (describing *Zarya of the Dawn* as an eighteen-page comic book written by Kashtanova with images generated using the Midjourney AI service).

[37] *See* Letter from Van Lindberg, Taylor English Duma LLP, to U.S. Copyright Office 2–4, 6–9 (Nov. 21, 2022) [hereinafter *Kashtanova Letter*] (arguing that Kashtanova's extensive prompting, selection, and refinement of Midjourney outputs rendered the images the product of her human authorship).

[38] *See Kasunic Letter*, *supra* note 36, at 1, 3–5 (recognizing Kashtanova as the author of the text and the selection, coordination, and arrangement of text and images, but stating that the Midjourney-generated images are "not the product of human authorship").

[39] *Id.* at 1, 4–5 (explaining that the new registration will have the same effective date as the original but will cover only the human-authored text and the compilation authorship in the selection, coordination, and arrangement of text and images, and will exclude the AI-generated images).

[40] Consistent with how many multi-modal AI (e.g., Google's Gemini) enable human users to engage with the AI in natural language while also possessing the ability to generate images. *See, e.g.*, *Generate & edit images with Gemini Apps*, GOOGLE, https://support.google.com/gemini/answer/14286560 [https://perma.cc/6ST9-T58V] (last visited Oct. 7, 2025); *Generate images with Gemini*, GOOGLE CLOUD, https://cloud.google.com/vertex-ai/generative-ai/docs/multimodal/image-generation [https://perma.cc/8JBV-TUWC] (last visited Oct. 7, 2025).

Kashtanova "conceived and executed" the images?[41] These and other novel questions become acute with fluid agency.

While the full legal implications of this unmappability are yet to be seen, preliminary evidence from disputes involving less advanced autonomous systems already demonstrates how this crisis of attribution is straining foundational legal doctrines. In the creative sphere, courts are confronting the entanglement of human direction and machine execution. A 2023 ruling by the Beijing Internet Court in *Li v. Liu*, for instance, granted copyright protection to an AI-generated image, reasoning that the plaintiff's detailed prompting and iterative adjustments constituted a sufficient "intellectual investment."[42] Yet in recognizing the human's role, the court's analysis underscored the fundamental difficulty of delineating where the user's creative guidance ended and the AI's autonomous synthesis began.[43]

---

[41] These questions implicate two distinct thresholds in current doctrine. The first is quantitative: how much input must a human provide? The second is qualitative: does that input constitute sufficient "creative control" over the work's expressive elements? The Copyright Office's 2023 guidance emphasizes that registration turns on "the extent to which the human had creative control over the work's expression," not merely the volume of prompts. *See* Copyright Registration Guidance: Works Containing Material Generated by Artificial Intelligence, 88 Fed. Reg. 16,190, 16,192 (Mar. 16, 2023), https://www.copyright.gov/ai/ai_policy_guidance.pdf [https://perma.cc/2L9R-L5JF] [hereinafter Copyright Registration Guidance]. The Office's 2025 Copyrightability Report reinforces this distinction, concluding that "prompts alone do not provide sufficient human control to make users of an AI system the authors of the output" because prompts convey unprotectable ideas rather than protectable expression. *See* U.S. COPYRIGHT OFFICE, COPYRIGHT AND ARTIFICIAL INTELLIGENCE PART 2: COPYRIGHTABILITY 18–19 (2025), https://www.copyright.gov/ai/Copyright-and-Artificial-Intelligence-Part-2-Copyrightability-Report.pdf [https://perma.cc/5HPV-7ETJ] [hereinafter U.S. Copyright Office, *Part 2: Copyrightability*]. Yet commenters caution that "exceptional" cases may exist where prompts are "so directive and detailed" that the user's expression is "directly perceptible in the machine's output." *Id.* at 22 n.80. This case-by-case approach leaves the precise application of these thresholds indeterminate—an ambiguity that fluid agency exacerbates.

[42] *See* Li Yunkai v. Liu Yuanchun, (2023) Jing 0491 Min Chu No. 11279 (Beijing Internet Ct. Nov. 27, 2023) (P.R.C.) (*English translation available at* https://patentlyo.com/media/2023/12/Li-v-Liu-Beijing-Internet-Court-20231127-with-English-Translation.pdf [https://perma.cc/VGV3-NZWR]).

This tension crystallized in litigation over *Théâtre D'opéra Spatial*, an AI-assisted image that won first prize at the 2022 Colorado State Fair.[44] After the U.S. Copyright Office issued a final refusal to register the work, its creator sued, arguing that his 624 prompt iterations and substantive post-production edits supplied the necessary human authorship.[45] The Copyright Office had refused to register the claim because the final image included "inextricably merged, inseparable contributions" from both human and machine.[46] On review, the Board found that the work contained more than a de minimis amount of AI-generated content that had to be disclaimed; because Allen refused to disclaim those AI-generated portions, the Office determined that the work could not be registered as submitted.[47]

A parallel challenge is unfolding in software development, exemplified by the ongoing class-action lawsuit against GitHub's Copilot.[48] Although framed as a copyright dispute, it prefigures the inventorship dilemma. Plaintiffs allege their code was unlawfully reproduced, while the Defendant argues that the AI's synthesis is so transformative that tracing any given output back to specific training data—the very act of attribution—is practically impossible.[49]

---

[43] *See id.* at 8-9 (discussing the court's evaluation of the plaintiff's intellectual input versus the AI's mechanical generation).

[44] *See* Complaint at 1–2, Allen v. Perlmutter, No. 1:24-cv-02665 (D. Colo. Sept. 26, 2024) https://storage.courtlistener.com/recap/gov.uscourts.cod.237436/gov.uscourts.cod.237436.1.0_1.pdf [https://perma.cc/Y2CM-7GRY] (describing the work's creation and its award at the Colorado State Fair).

[45] *Id.* at 8, 13 (describing Allen's iterative process and seeking judicial review); *see* also Letter from the U.S. Copyright Off. Rev. Bd., To Tamara Pester (Sept. 5, 2023), https://www.copyright.gov/rulings-filings/review-board/docs/Theatre-Dopera-Spatial.pdf [https://perma.cc/4F5X-ALQL] (noting Allen stated he "input numerous revisions and text prompts at least 624 times").

[46] *See* Letter from the U.S. Copyright Off. Rev. Bd., To Tamara Pester (Sept. 5, 2023), *supra* note 45, at 2.

[47] *Id.* at 1, 6 (finding "more than a de minimis amount of content generated by artificial intelligence" that must be disclaimed, and concluding that "[b]ecause Mr. Allen is unwilling to disclaim the AI-generated material, the Work cannot be registered as submitted").

[48] *See* Doe 1 v. GitHub, Inc., No. 4:22-cv-06823-JST (N.D. Cal. Nov. 3, 2022), https://storage.courtlistener.com/recap/gov.uscourts.cand.403220/gov.uscourts.cand.403220.1.0.pdf [https://perma.cc/R9RN-SL8E].

Similar crises of attribution are evident in tort and administrative law, where the locus of responsibility for AI-driven harm becomes fundamentally unmappable. The 2018 fatality involving an Uber self-driving vehicle in Tempe, Arizona provides a stark illustration. In that incident, a test vehicle operating in autonomous mode struck and killed a pedestrian after the system failed to correctly classify the hazard and the human safety driver failed to intervene. Subsequent investigation revealed an inextricable blend of the system's design limitations and the human safety driver's negligence.[50] Legal accountability, however, was ultimately deflected from the corporation to the human operator, demonstrating a "moral crumple zone"[51] where entangled causation defaults to the most proximate human actor.

---

[49] *Compare* Complaint at ¶¶ 2–4, *Doe 1*, No. 4:22-cv-06823-JST, https://storage.courtlistener.com/recap/gov.uscourts.cand.403220/gov.uscourts.cand.403220.1.0.pdf [https://perma.cc/R9RN-SL8E] (alleging that Copilot reproduces open-source code without license compliance), *with* Defendant GitHub, Inc.'s Motion to Dismiss at 1, 8, *Doe 1*, No. 4:22-cv-06823-JST (Dkt. 50), https://storage.courtlistener.com/recap/gov.uscourts.cand.403220/gov.uscourts.cand.403220.50.0.pdf [https://perma.cc/JCW3-23NT] (arguing that Copilot "generates its own code" based on what it has learned and that plaintiffs failed to identify any specific instances where their code was reproduced).

[50] *See* NAT'L TRANSP. SAFETY BD., COLLISION BETWEEN VEHICLE CONTROLLED BY DEVELOPMENTAL AUTOMATED DRIVING SYSTEM AND PEDESTRIAN, TEMPE, ARIZONA, MARCH 18, 2018 59 (2019), (concluding the probable cause was the safety driver's failure to monitor the driving environment, but citing Uber's inadequate safety risk assessment procedures and the system's inability to correctly classify the pedestrian as contributing factors).

[51] *See* Letter from Sheila Polk, Yavapai Cnty. Att'y, to Bill Montgomery, Maricopa Cnty. Att'y (Mar. 4, 2019) (declining to pursue criminal charges against Uber because there was "no basis for criminal liability" for the corporation); *see also* Madeleine Clare Elish, *Moral Crumple Zones: Cautionary Tales in Human–Robot Interaction*, 5 ENGAGING SCI. TECH. & SOC'Y 40 (2019) [https://doi.org/10.17351/ests2019.260] (coining the term "moral crumple zone" to describe how responsibility for a systemic failure is often misattributed to a human operator who had limited control).

The unmappability that creates "crumple zones" in physical torts extends to the erosion of procedural rights and the infliction of economic injury. In *Houston Federation of Teachers v. Houston Independent School District*, a federal court found that using a proprietary "value-added" algorithm to terminate teachers violated due process because its opaque and interdependent calculations made it impossible for educators to meaningfully challenge a potentially career-ending decision.[52] On a systemic scale, Australia's "Robodebt" scandal revealed a catastrophic failure of administrative accountability, where a hybrid automated-and-human system issued hundreds of thousands of unlawful welfare debt notices.[53] A subsequent Royal Commission concluded that the process was so entangled that officials could not reconstruct or justify how specific debts were generated, making it impossible to trace the causal pathway from individual decisions to specific harms.[54]

In each of these cases, the core issue remains the same: the generative process obscures provenance. Whether manifesting as an authorship dispute or a liability crisis, the complex interplay of human and machine action renders traditional tests of origin and contribution increasingly tenuous, straining legal frameworks designed to assign rights or fault to a discrete, identifiable source.

Having established the core challenge of unmappability, we now turn to its first major legal casualty: the doctrine of authorship, which, like other frameworks we will examine, presupposes the ability to attribute creative acts to specific, cognizable actors.

---

[52] *See* Houston Fed'n of Tchrs., Loc., Local 2415 v. Houston Indep. Sch. Dist., 251 F. Supp. 3d 1168, 1179 (S.D. Tex. 2017) (denying the school district's motion for summary judgment on procedural due process grounds because the opaque algorithm gave teachers no meaningful way to challenge a potentially career-ending decision).

[53] *See* ROYAL COMM'N INTO THE ROBODEBT SCHEME, REPORT (2023) (Austl.), https://robodebt.royalcommission.gov.au/system/files/2023-09/rrc-accessible-full-report.PDF [https://perma.cc/8DF9-2XP4] (finding that more than 400,000 unlawful debts were raised by the automated income-averaging system).

[54] *See id.* at xxiii–xxix (finding that the entanglement of "income averaging" algorithms and human processes created debts that the department could neither reconstruct and document nor legally substantiate).

# III. COPYRIGHT LAW

Scholarship has long questioned whether traditional copyright frameworks can accommodate works generated by algorithmic processes.[55] At the heart of the matter is a simple-to-state but difficult-to-answer question: when an AI generates intellectual content, *who is the author*—and thus *who owns the copyright*? The artist who commissions it? The provider who builds the system? The AI itself? Or perhaps no one at all?

U.S. copyright frameworks remain anthropocentric, hinging on whether a human has exercised meaningful control over AI-generated outputs.[56] The U.S. Copyright Office's 2023 policy affirms that AI-generated works lacking substantial human authorship cannot be copyrighted.[57] The Office applied this policy in its *Zarya of the Dawn* decision, where it refused to register the work's AI-generated images, holding that the user's prompts did not make her the "master mind" of the images and that the AI's output is not fully predictable.[58]

These human-centric paradigms face mounting criticism. Annemarie Bridy, for instance, challenges the entrenched assumption of uniquely human authorship by arguing creativity itself is inherently algorithmic.[59] She argues that even 'human' creativity operates through rules and structured processes, so works produced autonomously by computers are less alien to our creative paradigms than the law presumes.[60] If the law is to remain relevant, it must evolve to accommodate machine-generated creative content.

---

[55] For foundational scholarship, *see, e.g.*, Pamela Samuelson, *Allocating Ownership Rights in Computer-Generated Works*, 47 U. PITT. L. REV. 1185 (1986); Raquel Acosta, *Artificial Intelligence and Authorship Rights*, HARV. J.L. & TECH. DIG. (Feb. 17, 2012); Peter Jaszi, *Toward a Theory of Copyright: The Metamorphoses of "Authorship"*, 1991 DUKE L.J. 455; Ryan Abbott, *The Reasonable Robot: Artificial Intelligence and the Law*. CAMBRIDGE UNIV. PRESS (2020).

[56] *See* MARTIN ZEILINGER, TACTICAL ENTANGLEMENTS: AI ART, CREATIVE AGENCY, AND THE LIMITS OF INTELLECTUAL PROPERTY (Meson Press 2021).

[57] *See* Mark A. Lemley, *How Generative AI Turns Copyright Upside Down*, 25 COLUM. SCI. & TECH. L. REV. 190 (2024); Copyright Registration Guidance, *supra* note 41, at 16,192.

[58] *See* Kasunic Letter, *supra* note 36, at 7.

[59] *See* Bridy, *supra* note 12, at ¶ 2.

Bridy's argument, once forward-looking, now directly confronts a legal system grappling with the practical realities of these new creative paradigms. Rights-holders have been engaged in a legal battle on two fronts: the legality of using copyrighted works for AI training,[61] and the infringing nature of AI-generated outputs.[62] Chinese courts, in particular, have reached divergent conclusions. In *Beijing Film Law Firm v. Baidu*, the Beijing Internet Court held that content autonomously generated by software is not a "work," reasoning that Chinese copyright protects only the intellectual creations of natural persons.[63] By contrast, in *Shenzhen Tencent*, the Shenzhen Nanshan District People's Court granted copyright protection to an AI-generated news article. The court emphasized human choices in preparing and organizing data inputs, setting trigger conditions, templates, and corpora, and training a proofreading algorithm.[64] These activities, while less direct than the prompting in *Zarya*,[65] were sufficient to support protection under Chinese law. This divergence highlights a fundamental tension: Is direct, expressive input (like detailed prompting) the *sine qua non* of authorship, or can more indirect, preparatory contributions suffice?

---

[60] *Id*. at 27.

[61] *See* Complaint, N.Y. Times Co. v. Microsoft Corp. & OpenAI, Inc., No. 1:23-cv-11195 (S.D.N.Y. Dec. 27, 2023) (news articles); Complaint, Kadrey v. Meta Platforms, Inc., No. 3:23-cv-03417 (N.D. Cal. July 7, 2023) (books); Complaint, Andersen v. Stability AI Ltd., No. 3:23-cv-00201 (N.D. Cal. Jan. 13, 2023) (visual art); Complaint, Getty Images (US), Inc. v. Stability AI, Inc., No. 1:23-cv-00135 (D. Del. Feb. 3, 2023) (photographs); Complaint, UMG Recordings, Inc. v. Suno, Inc., No. 1:24-cv-11611 (D. Mass. June 24, 2024) (sound recordings); *see also* Thomson Reuters Enter. Ctr. GmbH v. ROSS Intelligence Inc., No. 20-613-SB, 2025 WL 458520 (D. Del. Feb. 11, 2025) (granting partial summary judgment on direct infringement and rejecting a fair-use defense).

[62] As exemplified by the suit from major film studios against Midjourney. *See* Complaint at 2, Disney Enters., Inc. v. Midjourney, Inc., No. 2:25-cv-05275 (C.D. Cal. June 11, 2025) (alleging "Midjourney is the quintessential copyright free-rider and a bottomless pit of plagiarism"). The studios alleged direct and secondary copyright infringement based on both the unauthorized copying of works to train the AI model and the subsequent generation of outputs substantially similar to iconic characters. *Id.*at 101-107.

[63] Beijing Film Law Firm v. Beijing Baidu Netcom Sci. & Tech. Co., China Daily, (Beijing Internet Ct. Apr. 25, 2019) (China) (holding that only works created by natural persons qualify for copyright protection).

[64] *See* Shenzhen Tencent Comput. Sys. Co. v. Shanghai Yingxun Tech. Co., China Justice Observer (Shenzhen Nanshan Dist. People's Ct. Dec. 24, 2019) (China) (granting copyright to an AI-generated article based on human involvement in selecting and arranging inputs). For a detailed comparison, *see* Yin Wan & Hui Lu, *Copyright Protection for AI-Generated*

## A. The Unworkability of Contribution-Based Frameworks

Crucially, this question, key debates concerning authorless versus authored works,[66] and proposed solutions such as the UK's computer-generated works approach that vests authorship in the person making the necessary arrangements,[67] two-tiered protection systems,[68] and Gervais's theory of originality causation,[69] all turn on *parsing* human and AI contributions. It is only *if* human and AI contributions can be clearly delineated that we can credit a human author for creative direction and either acknowledge the AI's role in a new category (e.g., "AI-assisted creation") or attribute the AI-generated portions to the human by extension.[70] Or we can shift the inquiry from status to weight: instead of asking "who is the author?", ask "how much did each contribute?" A work in which contributions can be separated, for example, could trigger a royalty allocation among the commissioning user, the model developer, and a fund for creators

---

*Outputs: The Experience from China*, 42 COMPUT. L. & SEC. REV. 105581 (2021) [https://doi.org/10.1016/j.clsr.2021.105581].

[65] *Cf.* Kasunic Letter, *supra* note 36, at 7 (describing the user's process of entering text prompts to generate images).

[66] *See* Lemley, *supra* note 57, at 192-194.

[67] *See* Ryan Abbott, *Artificial Intelligence, Big Data and Intellectual Property: Protecting Computer Generated Works in the United Kingdom*, *in* RESEARCH HANDBOOK ON INTELLECTUAL PROPERTY AND DIGITAL TECHNOLOGIES 322–37 (Tanya Aplin ed. 2020); *see also* Copyright, Designs and Patents Act 1988, c. 48, §§ 9(3), 178 (UK).

[68] *See* Haochen Sun, *Redesigning Copyright Protection in the Era of Artificial Intelligence*, 107 IOWA L. REV. 1213 (2022).

[69] *See* Daniel J. Gervais, *The Machine as Author*, 105 IOWA L. REV. 2053, 2099–2101 (2020).

[70] Whether this is advisable is another question, with arguments falling on both sides. *Compare* James Grimmelmann, *There's No Such Thing as a Computer-Authored Work—and It's a Good Thing, Too*, 39 COLUM. J.L. & ARTS 403 (2016) (arguing against AI authorship), *with* Sun, *supra* note 68 (proposing *sui generis* rights for AI-generated works with human inputs).

whose works trained the AI.[71] Or we can consider *sui generis* rights—limited protections weaker than full human authorship but stronger than the public domain.[72]

These questions, debates, and proposed solutions become largely insoluble if human-AI contributions cannot be parsed. A framework premised on distinguishing the origin of creative elements faces two intractable problems: (1) ensuring fair and consistent treatment across cases where contributions are separable versus those where they are inseparable, and (2) establishing reliable criteria for determining whether contributions can even be parsed in the first place.

Consider two classes of works resulting from human-AI interaction. Works in the first (separable) class allow specific creative elements to be reasonably attributed to either the human or the AI. For example, the human might have written distinct sections while an AI generated others. In the second (inseparable) class, the interaction, likely involving recursive feedback loops, results in an inextricably blended work—a fusion where the origins of specific ideas, phrasings, or creative choices are fundamentally entangled and untraceable.[73]

---

[71] For example, a human who heavily edited the AI's output might receive a larger share, while a largely AI-generated work might favor the developer. Any such calibration, however, lives or dies by proxies (e.g., edit counts, prompt iterations, selection logs), which are themselves contestable and susceptible to gaming. Notably, recent European legislation creates transparency and record-keeping hooks while stopping short of creating a direct royalty framework for AI outputs. The EU AI Act requires event logging for *high-risk* systems (Art. 12) and obliges providers of general-purpose AI models to adopt a policy to comply with Union copyright law, including honoring reservations of rights under DSM Directive art. 4(3) (Art. 53(1)(c)). The EU Data Act permits reasonable, non-discriminatory compensation for making data available (Art. 9), but this does not create AI output-royalty entitlements. *See* Regulation (EU) 2024/1689 of the European Parliament and of the Council of 13 June 2024 laying down harmonised rules on artificial intelligence (Artificial Intelligence Act), OJ L, 2024/1689, 12.7.2024; Regulation (EU) 2023/2854 of the European Parliament and of the Council of 13 December 2023 on harmonised rules on fair access to and use of data (Data Act), O.J. (L 2023/2854), 22.12.2023; *see also*, e.g., João Pedro Quintais, *Generative AI, Copyright and the AI Act*, 56 COMPUT. L. & SEC. REV. 106107 (2025).

[72] Unlike proportional royalties, which operate within existing copyright frameworks to distribute revenue, *sui generis* rights create a new framework with its own rules for protection, duration, and scope. Existing *sui generis* regimes like the EU Database Directive, protecting non-creative investments (e.g., data compilation) for 15 years, may offer a precedent for AI-generated works. *See* Directive 96/9/EC of the European Parliament and of the Council of 11 March 1996 on the legal protection of databases, 1996 O.J. (L 77) 20; *see also* J.H. Reichman & Pamela Samuelson, *Intellectual Property Rights in Data*, 50 VAND. L. REV. 51, 52 (1997). This approach sidesteps the authorship-attribution problem but risks incentivizing a flood of AI-generated content that could depress the value of human-created works.

This division yields a dilemma: no single legal standard can treat both classes coherently. On the one hand, a framework *based on* separating human and AI contributions (e.g., granting full copyright only to human-generated portions) collapses when applied to the inseparable class, because the necessary distinctions cannot be drawn. On the other hand, a framework suitable for inseparable works must operate *without* assessing the extent of specific contributions. Yet if applied to the separable class, that same framework is too blunt, treating works with vastly different contribution levels identically.

The obvious rejoinder is a two-track regime: one standard for separable works and another for inseparable ones (or an inseparability exception layered onto existing rules). But that simply relocates the problem to a contested threshold: whether a work is truly separable. This determination is subjective and prone to inconsistent line-drawing along the human-AI continuum. Mixed cases aggravate the puzzle: does any inseparable element "contaminate" the whole, pushing it into the inseparable track, or can parties slice a work into putatively separable fragments to force the other track? Either approach invites over- or under-inclusion. Unless the two standards are calibrated to converge in *outcomes*, similarly situated works will receive different legal treatment based on the happenstance of traceability rather than the substance of human effort.

---

[73] The challenge is analogous to the difficulty of separating protectable expression from unprotectable function in computer software, which led to the "Abstraction–Filtration–Comparison" test. *See* Computer Assocs. Int'l, Inc. v. Altai, Inc., 982 F.2d 693, 706–11 (2d Cir. 1992); *see also* Gates Rubber Co. v. Bando Chem. Indus., 9 F.3d 823, 834–46 (10th Cir. 1993) (adopting and elaborating AFC). In practice, courts have struggled to police that boundary where expressive choices are intertwined with functionality. *Cf.* Lotus Dev. Corp. v. Borland Int'l, Inc., 49 F.3d 807, 813–19 (1st Cir. 1995) (holding a menu command hierarchy to be an uncopyrightable "method of operation" under § 102(b)), *aff'd by an equally divided Court*, 516 U.S. 233 (1996); Oracle Am., Inc. v. Google Inc., 750 F.3d 1339, 1354–61 (Fed. Cir. 2014) (holding API declaring code copyrightable), *but see* Google LLC v. Oracle Am., Inc., 141 S. Ct. 1183 (2021) (assuming without deciding that the declaring code was copyrightable and resolving the case on fair use grounds, leaving copyrightability unresolved).

## B. Recursive Adaptation

Several novel challenges follow from the co-evolutionary, recursive adaptation described in Part I—where an AI adapts to its human user's prompts and feedback and the human user adjusts to the AI's responses—that make any principle based on differentially treating a work by source (AI-generated versus human-authored) infeasible.[74]

Consider a large language model (e.g., ChatGPT) that evolves to align with its human user's preferences, absorbing and fine-tuning its outputs based on the user's inputs and feedback. Simultaneously, the human user evolves her style and knowledge base, learning from the AI.[75] When neither the human user nor the AI fully determines the creative trajectory and the AI system itself becomes an active participant in the evolution of a human user's style, how should authorship rights be apportioned?

Suppose the AI's output satisfies *Feist*'s "modicum of creativity" standard.[76] Under current U.S. law, copyright protection turns on human authorship, not originality alone.[77] Therefore, this output has no author.

However, could its human user nonetheless assert a copyright claim on the output as a derivative work of her own pre-existing copyrighted works?[78] By the process of recursive adaptation, if the AI's output is *functionally derivative*[79]—a recombination rather than a transformation of the user's prior copyrighted expressions that had been shared with the AI—should the fact that the output was from an AI matter?

---

[74] These issues are distinct from debates surrounding traditional generative AI, which primarily focus on AI's training and whether its outputs are substantially similar to the training data. *See* Weijie Huang & Xi Chen, *Does Generative AI Copy? Rethinking the Right to Copy Under Copyright Law*, 56 COMPUT. L. & SEC. REV. 106100 (2025). Here, in contrast, we raise questions regarding the similarity of AI outputs to user inputs, post-training.

[75] Linguistic convergence is a well-established phenomenon whereby conversational partners begin to mimic each other's language use, such as word choices, phrasings, and syntactic structures. *See* Pickering & Garrod, *supra* note 30. In human-AI interactions, users may similarly adapt to an AI's language use over time. *See, e.g.*, Vinchon et al., *supra* note 30. For the technical grounding of user-specific adaptation, *see* Zhang, *supra* note 26.

[76] Feist Publ'ns, Inc. v. Rural Tel. Serv. Co., 499 U.S. 340, 345 (1991).

[77] *See* Thaler v. Perlmutter, 130 F.4th 1039, 1041 (D.C. Cir. 2025) (holding that, "[a]s a matter of statutory law, the Copyright Act requires all work to be authored in the first instance by a human being"); U.S. Copyright Office, *Part 2: Copyrightability*, *supra* note 41, at 8–16 & Exec. Summ. iii; Copyright Registration Guidance, *supra* note 41, at 16,192.

[78] *See* 17 U.S.C. § 101 (2018) (defining "derivative work"); L. Batlin & Son, Inc. v. Snyder, 536 F.2d 486, 491–92 (2d Cir. 1976) (*en banc*) (derivative work must add more than trivial variation); Meshwerks, Inc. v. Toyota Motor Sales U.S.A., 528 F.3d 1258, 1265–69 (10th Cir.

Present doctrine makes this path vanishingly narrow. Even if the AI's output merely recombines the user's own pre-existing, copyrighted expression, the AI-generated portions are uncopyrightable, and protection, if any, extends only to the perceptible human-authored contributions in the final work.[80] As a practical matter, that leaves the user unable to secure protection for the new AI-mediated recasting of her prior expression, even when the recombination reflects the user's style through the AI's adaptations.

---

2008) (faithful digital reproduction lacks the necessary original authorship); Schrock v. Learning Curve Int'l, Inc., 586 F.3d 513, 519–22 (7th Cir. 2009) (explaining that an authorized derivative work is copyrightable if the author's incremental contributions are original).

[79] Suppose AI incorporates elements from a human user's inputs verbatim into its outputs. *See, e.g.*, Bender et al., *supra* note 34, at 617. The AI's outputs then might even be better characterized as a curation of its human user's creativity, rather than as independent creation. This raises the possibility that, even though an AI lacks the *mens rea* associated with copyright authorship and a work generated without human authorship is not registrable under current U.S. law, the AI's outputs could be considered *functionally* derivative of its human user's inputs. *See* 17 U.S.C. § 101 (2018); Copyright Registration Guidance, *supra* note 41, at 16,192. Moreover, allegations that models can memorize and reproduce training data with high fidelity lend weight to this argument, blurring the line between generation and sophisticated recombination. *See* Milad Nasr, Nicholas Carlini, Jonathan Hayase, Matthew Jagielski, A. Feder Cooper, Daphne Ippolito, Christopher A. Choquette-Choo, Eric Wallace, Florian Tramèr & Katherine Lee, *Scalable Extraction of Training Data from (Production) Language Models*, ARXIV (Nov. 28, 2023), https://arxiv.org/pdf/2311.17035 [https://perma.cc/CK2H-52Q8] (demonstrating extractable memorization from production models); *see also* Amended Complaint ¶ 253, Authors Guild v. OpenAI Inc., No. 1:23-cv-08292 (S.D.N.Y. Dec. 4, 2023) (alleging that "When prompted, ChatGPT accurately generated summaries of several of the Martin Infringed Works…").

[80] *See* 17 U.S.C. § 103(b) (2018) (copyright in a derivative work extends only to the material contributed by the author); U.S. Copyright Office, *Part 2: Copyrightability*, *supra* note 41, at 10, 24–31 (summarizing the requirement to disclose AI-generated material and discussing modifications to AI outputs); Copyright Registration Guidance, *supra* note 41, at 16,193.

The dilemma is compounded when considering any *novel* elements in the output. For these new contributions to be protected, the user bears the burden of proving they originated from human authorship, not the AI's autonomous generation. Yet, as a direct consequence of fluid agency and recursive adaptation, the provenance of any given creative element is often irreducibly ambiguous. In the face of this unmappability, the legal default prevails: where the origin of a novel element cannot be clearly attributed to the human user, it is treated as uncopyrightable machine generation.[81] The user is thus trapped: unable to protect the AI-mediated recasting of old expression in its new form, and unable to claim ownership of new expression whose human origins cannot be definitively proven.

This holds true even if the final output is functionally transformative, akin to a collage artist transforming source material.[82] Suppose a key element emerges entirely from the AI, but there is strong evidence that the AI assimilated the user's style and prior works. Because purely AI-generated novel contributions are uncopyrightable, those elements are unprotected (while any human-authored portions of the work remain protected), even if they predictably reflect the user's inputs or style.[83]

Even a hypothetical legal fix—for instance, allowing users to assert copyright in functionally derivative AI outputs—merely inverts the problem. Under the current regime, an AI's output is presumed to be uncopyrightable machine generation; a user bears the burden of proving her own creative contribution to claim any rights.[84] Under the hypothetical fix, an AI's output could be presumed to be a derivative work owned by its human user. A challenger would then bear the burden of proving that a creative element was *not* derivative—that it was, in fact, an independent product of the AI. In a system defined by fluid agency, however, each output is a probabilistic function of an AI's entire interaction history with the user.[85] Proving a specific creative idea was "independent" requires unscrambling the causal egg: tracing the idea back through a recursively adaptive feedback loop to show that it had no connection to the user's preferences and inputs. This task is both technically and conceptually infeasible, reintroducing the unmappability problem from a different angle.

---

[81] Under current Office practice, applicants must be able to identify perceptible human-authored contributions; absent such evidence, contested material is treated as unprotectable machine generation. *See* U.S. Copyright Office, *Part 2: Copyrightability*, *supra* note 41, at Exec. Summ. iii & 24–31.

[82] *See* Campbell v. Acuff-Rose Music, Inc., 510 U.S. 569, 579 (1994); Cariou v. Prince, 714 F.3d 694, 706–09 (2d Cir. 2013); *see also* Andy Warhol Found. for the Visual Arts, Inc. v. Goldsmith, 598 U.S. 508, 523–35 (2023) (first-factor analysis focuses on the specific use's purpose; 'new meaning or message' alone is not dispositive). *But cf.* 17 U.S.C. § 101 (2018) (defining a "derivative work" as one that is "recast, transformed, or adapted*");* Micro Star v. FormGen Inc., 154 F.3d 1107, 1112–16 (9th Cir. 1998) (distinguishing fair use transformation from the creation of a derivative work).

[83] *See* U.S. Copyright Office, *Part 2: Copyrightability*, *supra* note 41, at Exec. Summ. iii & 24–31.

[84] *See supra* note 81.

These issues are novel because until recently, AI lacked the capacity for long-term memory and significant adaptation. With earlier systems, it sufficed to consider a human user's prompts and curation to judge contribution.[86] With fluid agency, however, denying AI outputs authorship *carte blanche* becomes tantamount to denying human users authorship when they express their ideas through AI. Yet enabling such claims remains impracticable, as it requires determining if users "exercise[d] creative control over an AI's output and contribute[d] original expression" through iterative refinements—a standard that becomes unworkable in complex, multi-turn interactions.[87] The result is a doctrinal mismatch where origin-based tests presuppose a separability that recursive human-AI feedback loops effectively dissolve.

## C. Other Foundational Doctrines

The unmappability engendered by fluid agency not only challenges the core definition of authorship but also destabilizes adjacent doctrines that are similarly predicated on a discernible human creator. The work-made-for-hire, joint authorship, and moral rights doctrines each falter when confronted with irreducibly entangled human-machine contributions.

First, the work-made-for-hire (WMFH) doctrine treats the employer as the author *ab initio*, vesting initial ownership in a work prepared by an employee within the scope of employment or in certain specially commissioned works under a signed agreement.[88] This doctrine, however, presupposes a human creator—a premise shared across many legal systems, despite their different approaches to initial ownership.[89] If an AI's autonomy is so significant that no human qualifies as the legal "author," the WMFH doctrine cannot attach, because there is no underlying "work of authorship" for ownership to vest in the employer.[90] This then forces courts into an unworkable inquiry: determining whether a human employee provided "sufficient creative direction" over the AI's process to be deemed the author. Such a standard, much like the "superintendence" test for joint authorship, would be difficult to apply when a non-human agent's contribution is substantial and inextricably merged with a human's.[91]

---

[85] *See supra* Part I (defining fluid agency and describing its stochastic, dynamic, and adaptive nature).

[86] *See* Kasunic Letter, *supra* note 36, at 7.

[87] *See* U.S. Copyright Office, *Part 2: Copyrightability*, *supra* note 41, at 24–31; Copyright Registration Guidance, *supra* note 41, at 16,192.

[88] 17 U.S.C. §§ 101, 201(b) (2018) (defining a "work made for hire" and vesting ownership in the employer); *see also* Cmty. for Creative Non-Violence v. Reid, 490 U.S. 730, 751–52 (1989) (establishing that "employee" status is determined under common law agency

Second, joint authorship standards are inapplicable when one contributor is an AI. Under U.S. law, a "joint work" is statutorily defined as one "prepared by two or more authors with the intention that their contributions be merged into a unitary whole."[92] An AI

---

factors); RESTATEMENT (THIRD) OF AGENCY § 7.07 (AM. L. INST. 2006) (defining scope of employment).

[89] The U.S. 'work-made-for-hire' doctrine automatically vests copyright ownership in the employer. By contrast, many civil-law systems rooted in *droit d'auteur* initially vest rights in the employee-creator. *See, e.g.*, Urheberrechtsgesetz [UrhG] [Copyright Act], Sep. 9, 1965, BGBl. I at 1273 § 43 (Ger.); Code de la propriété intellectuelle [CPI] [Intellectual Property Code] art. L111-1 (Fr.) (general rule of author-vesting), *with statutory exceptions for software*, *e.g.*, UrhG § 69b (Ger.); CPI art. L113-9 (Fr.). The UK's statutory approach differs; under the Copyright, Designs and Patents Act 1988, c. 48 § 11(2) (UK), the employer is generally the first owner of copyright in an employee's work. Notably, the UK also provides a rule for "computer-generated" works, assigning authorship to "the person by whom the arrangements necessary for the creation of the work are undertaken," *id.* § 9(3)—a pragmatic legal construct to fill the attribution gap when a traditional human author is absent. For a comparative discussion, see Dániel Legeza, *Employer as Copyright Owner from a European Perspective* (paper presented at the SERCI Annual Congress, 2015).

[90] *See* 17 U.S.C. § 102(a) (2018) (stating that copyright protection subsists in "original works of authorship"); *Thaler*, *supra* note 77, at 1041; Copyright Registration Guidance, *supra* note 41, at 16,190.

[91] *See* Aalmuhammed v. Lee, 202 F.3d 1227, 1234–35 (9th Cir. 2000) (discussing "superintendence" in the context of determining joint authorship).

[92] 17 U.S.C. § 101 (2018). Courts have distilled this into a two-part test requiring that each putative co-author (1) intend to be a co-author and (2) contribute independently copyrightable expression. *See, e.g.*, Childress v. Taylor, 945 F.2d 500, 507–08 (2d Cir.

can satisfy neither prong: as a non-person, neither can it supply legally cognizable copyrightable expression nor can it form the requisite intent.[93] While proving subjective intent among human collaborators is a familiar evidentiary challenge,[94] AI transforms this problem from one that is merely evidentiary to one that is fundamentally ontological. An AI, lacking legal personhood and consciousness, has no "prior state of mind" to probe and no intent to infer. Even a perfect log of the recursive interaction documents a process of mechanistic responses to human queries. The joint authorship doctrine is therefore fundamentally inapplicable.

Third, moral rights—such as the right to attribution and the right to integrity of a work—are inherently tied to a human author's personal connection to their creation.[95] While a cornerstone of copyright in many civil law jurisdictions, these rights receive only minimal protection in the United States.[96] AI, lacking legal personhood, cannot hold moral rights, but the unmappability engendered by fluid agency challenges the protection of these rights for the human user. For instance, if an AI modifies a work in ways that diverge from the human user's original artistic vision, the user's right to the *integrity* of the work may be challenged. Consider an AI literary agent that autonomously revises a manuscript

---

1991); Thomson v. Larson, 147 F.3d 195, 199–205 (2d Cir. 1998); Erickson v. Trinity Theatre, Inc., 13 F.3d 1061, 1068–71 (7th Cir. 1994).

[93] Joint authorship in European copyright law also generally requires a collaborative effort and a shared intention to create a unified work, presupposing human co-authors. *See, e.g.*, Urheberrechtsgesetz [UrhG] [Copyright Act] § 8 (Ger.); Code de la propriété intellectuelle [CPI] [Intellectual Property Code] art. L113-2 (Fr.); Copyright, Designs and Patents Act 1988, c. 48, § 10 (UK).

[94] As the Second Circuit noted in *Childress*, while the intent "at the time the writing is done" remains the "touchstone," "subsequent conduct is normally probative of a prior state of mind." *Childress*, 945 F.2d at 508–09.

[95] Moral rights are a cornerstone of copyright law in many civil law jurisdictions, stemming from the Berne Convention for the Protection of Literary and Artistic Works art. 6bis, Sept. 9, 1886, *as revised at* Stockholm, July 14, 1967, 25 U.S.T. 1341, 828 U.N.T.S. 221. *See, e.g.*, Urheberrechtsgesetz [UrhG] [Copyright Act] §§ 12–14 (Ger.); Code de la propriété intellectuelle [CPI] [Intellectual Property Code] art. L121-1 (Fr.).

[96] In the U.S., protection is limited primarily to certain works of visual art under the Visual Artists Rights Act of 1990 (VARA), 17 U.S.C. § 106A (2018). Courts have also rejected attempts to use trademark law to create a general right of attribution. *See* Dastar Corp. v. Twentieth Century Fox Film Corp., 539 U.S. 23, 33–34 (2003).

to incorporate themes its human author never endorsed. While the AI's alterations might even enhance the work, they undermine the author's right to control expression. Unlike traditional scenarios where moral rights protect against derogatory treatment by other humans, here an AI—employed by the human user—autonomously alters the work, reflecting a novel conflict between AI agency and user control.

# IV. PATENT LAW

The issues challenging authorship also arise in inventorship. A case in point is the *DABUS* litigation, which involves an AI system that generates novel inventions.[97] Patent applications naming *DABUS* as the *inventor* have been filed in the U.S., Europe, and the U.K., among other jurisdictions. Thus far, patent offices and courts in these jurisdictions have rejected these claims, insisting that inventors must be natural persons.[98]

---

[97] "DABUS" (Device for the Autonomous Bootstrapping of Unified Sentience) is an AI "creativity machine" developed by Stephen Thaler. It has been described as a system of neural networks in which one network generates variations on its training data and another evaluates those outputs for novelty and usefulness, and has reportedly produced concepts including a fractal-walled food container and a flashing emergency beacon for attracting attention in emergencies. See World Intellectual Property Organization [WIPO] Standing Comm. on the Law of Patents, *Artificial Intelligence and Inventorship*, SCP/35/7 Annex ¶¶ 114–16 (Sept. 13, 2023) (summarizing the operation of DABUS and its alleged inventions); Ryan Abbott, *Intellectual Property and Artificial Intelligence: An Introduction*, in RESEARCH HANDBOOK ON INTELLECTUAL PROPERTY AND ARTIFICIAL INTELLIGENCE 2, 11–13, 17 (Ryan Abbott ed., 2022) (describing DABUS as an AI "creativity machine" and explaining its two-network architecture); Ryan Abbott, *The Artificial Inventor Project*, WIPO MAGAZINE, (Dec.2019), https://www.wipo.int/wipo_magazine/en/2019/06/article_0002.html (overviewing the Artificial Inventor Project and its use of DABUS as the named inventor); David Yi, *AI Inventorship on the Horizon: Part 1*, NORTON ROSE FULBRIGHT (Oct. 2021), https://www.nortonrosefulbright.com/en/knowledge/publications/2a3c551a/ai-inventorship-on-the-horizon-part-1 [https://perma.cc/ZL8G-DCSL] (providing a practitioner-oriented description of DABUS and the beverage container and flashing light inventions).

[98] *See, e.g.*, Thaler v. Vidal, 43 F.4th 1207, 1213 (Fed. Cir. 2022), *cert. denied*, 143 S. Ct. 1783 (2023) (affirming the USPTO's rejection); Decision on Petition, *In re Application of Stephen L. Thaler*, Application No. 16/524,350 (U.S. Patent and Trademark Office Apr. 22, 2020); EPO Legal Bd. of Appeal, Decision J 8/20 (Dec. 21, 2021); Thaler v. The Comptroller-General of Patents, Designs and Trade Marks, [2021] EWCA Civ 1374 (Eng.), *aff'd*, [2023] UKSC 49; Comm'r of Patents v. Thaler, [2022] FCAFC 62, 289 FCR 45 (Austl.) (overturning

The legal questions in the *DABUS* case were relatively clear-cut only because no human participated in the inventive process. How might the outcome have differed if a human played some role, however minor, in ideation or development? One could imagine a continuum from no human participation to solely human participation, with AI systems potentially being fine-tuned or development processes adjusted to facilitate a human-AI partnership anywhere along that spectrum. At what point along the continuum would the legal standard for inventorship have been met?[99] And critically, would human-AI contributions be separable at that juncture, such that a standard based on contribution would be practicable?[100]

Specifically, under U.S. patent law, bringing an invention to fruition requires both *conception* ("the complete performance of the mental part of the inventive act") and *reduction to practice* (embodying the invention in a tangible form).[101] Inventorship turns on conception, and courts have long held that only humans can conceive inventions, meaning only natural persons can be legally recognized as inventors.[102]

A modern AI, however, may *functionally conceive* an innovation by generating novel solutions that would otherwise meet patentability criteria such as non-obviousness and utility if the inventor were human. For example, an AI drug discovery system might simulate new molecular structures addressing a target disease mechanism. The human-only conception requirement depends on identifying the extent to which the invention drew upon human user inputs and feedback. As seen in the *DABUS* case, if it is clear that the AI performed the core conception, the invention will lack a legally valid conceiver. However, if a human and an AI recursively adapt to one another, it may be unclear where a conception

---

Thaler v. Comm'r of Patents, [2021] FCA 879 (Austl.)). South Africa, a non-examining jurisdiction, granted a patent naming DABUS as inventor. *See* Companies & Intellectual Property Comm'n, S. Afr., Patent Application No. 2021/03242, *Notice of Acceptance* (June 24, 2021).

[99] For example, contrast the varying decisions of the Chinese courts as discussed earlier with the *DABUS* case. *See supra* notes 42, 63 and 64.

[100] At this threshold, can contributions be measured reliably enough to apply the standard?

[101] Burroughs Wellcome Co. v. Barr Labs., Inc., 40 F.3d 1223, 1227–28 (Fed. Cir. 1994).

[102] This issue extends to other jurisdictions. For instance, under the European Patent Convention (EPC), an inventor must be a natural person. *See* EPO Legal Bd. of Appeal, J 8/20 (Dec. 21, 2021) (holding that the EPC requires an inventor to be a natural person with legal capacity); *see also* EPC art. 81 (requiring the designation of the inventor); EPC r. 19(1) (specifying the content of the designation).

originated, obscuring whether the human or the AI performed the crucial "mental part of the inventive act."

The challenge extends to the second prong of inventorship: reduction to practice. This can be achieved either by physically embodying the invention and demonstrating its utility (*actual* reduction to practice) or by filing a patent application that satisfies the disclosure requirements of 35 U.S.C. § 112(a), thereby effecting *constructive* reduction to practice.[103] AI complicates both pathways.

For actual reduction to practice, AI systems integrated with robotics or simulation tools can perform the necessary physical steps or conduct *in silico* testing—though simulations alone typically do not establish actual reduction to practice for composition or device claims.[104] For instance, an AI might design, synthesize, and test a novel compound without direct human intervention in each step. However, when the AI executes these tasks—blending its own learned strategies with human inputs—attributing a successful reduction to practice becomes legally tenuous. The critical question is not *who performed the work*, but whether the named human inventor contributed to its conception, not merely its execution.

The hurdles are perhaps even higher for constructive reduction to practice. While AI can generate detailed technical descriptions, the enablement and written description requirements of § 112(a) pose distinct hurdles. Enablement demands that the disclosure teach a Person Having Ordinary Skill in the Art (PHOSITA) how to make and use the invention across the full scope of the claims without undue experimentation.[105] If the AI's inventive process relies on logic opaque to humans, the specification may describe a successful outcome but fail to disclose the operative steps or parameters a PHOSITA would need for replication, risking non-enablement.[106] The written description requirement shifts the inquiry from what the disclosure teaches the public to what the human inventor possessed at the time of filing.[107] When an AI is central to conception, can a human be said to truly possess the invention? The use of AI clouds the evidentiary link and challenges the conceptual link between inventor and invention.

---

[103] *See, e.g.*, Hyatt v. Boone, 146 F.3d 1348, 1352 (Fed. Cir. 1998) (stating that constructive reduction to practice occurs when an inventor files a patent application that complies with § 112).

[104] *See, e.g.*, Scott v. Finney, 34 F.3d 1058, 1061–62 (Fed. Cir. 1994) (actual reduction to practice requires a demonstration that the invention works for its intended purpose); Cooper v. Goldfarb, 154 F.3d 1321, 1327 (Fed. Cir. 1998).

[105] *See* Amgen Inc. v. Sanofi, 598 U.S. 594, 610–11 (2023).

[106] *See In re* Wands, 858 F.2d 731, 737 (Fed. Cir. 1988) (listing factors for determining undue experimentation). As an analogy, courts have shown skepticism toward "black box" disclosures in other contexts; for instance, computer-implemented means-plus-function claims are held indefinite under § 112(b) for failure to disclose a corresponding algorithm.

Moreover, similar to the challenges identified in authorship, the doctrine of joint inventorship faces distinct and novel difficulties when confronted with AI. Under current U.S. patent law, joint inventors must each contribute to the invention's conception—"the complete performance of the mental part of the inventive act"—and typically engage in some form of collaborative activity.[108] AI disrupts this framework by introducing a non-human entity capable of independently generating inventive concepts, yet incapable of forming the requisite collaborative connection or holding legal status as an inventor. Consider the prior example from drug discovery: An AI system, guided by human researchers, autonomously identifies a novel molecular structure constituting the core inventive concept. The AI's contribution meets technical criteria (novelty, utility), but neither can it form a collaborative relationship with a human, nor can it be named an inventor.

Can the human researchers be named? If a *single* researcher provided only high-level objectives, her contribution might fail the conception standard. If *multiple* researchers provided detailed specifications and iterative feedback, their collective contribution may be stronger—yet they still may not have conceived the specific, critical insight generated by the AI.

This presents a dilemma: How should inventorship be determined? If the AI is viewed simply as a sophisticated tool, perhaps the human researcher(s) should receive full

---

*See, e.g.*, Aristocrat Techs. Austl. Pty Ltd. v. Int'l Game Tech., 521 F.3d 1328, 1333 (Fed. Cir. 2008).

[107] *See* Ariad Pharms., Inc. v. Eli Lilly & Co., 598 F.3d 1336, 1341 (Fed. Cir. 2010) (en banc) (holding that the written description requirement is separate and distinct from the enablement requirement).

[108] *See* Burroughs Wellcome, *supra* note 101, at 1227–28 (defining conception); Ethicon, Inc. v. U.S. Surgical Corp., 135 F.3d 1456, 1460 (Fed. Cir. 1998) (requiring each joint inventor to contribute to conception); *see also* Kimberly-Clark Corp. v. Procter & Gamble Distrib. Co., 973 F.2d 911, 917 (Fed. Cir. 1992) (indicating joint inventors usually collaborate or show connection). While the standard for collaboration in U.S. patent law differs from the intent requirement articulated in Childress, *supra* note 94, the requirement for some joint effort remains. European frameworks generally concur. *See* EPC, *supra* note 102, art. 60.

inventorship credit, regardless of whether their contribution met traditional conception standards for the *entire* invention. If the principle from the *DABUS* litigation is applied strictly to the conception of the core inventive step, then perhaps no valid human inventor exists for that crucial AI-generated insight, potentially jeopardizing patentability even with significant human involvement. The challenge is compounded by fluid agency: was its critical insight truly independent, or was it functionally derived from the dynamic and adaptive feedback loops of prior human inputs? If traceable, did the insight arise primarily from the AI's adaptations to one specific researcher's inputs, or did it reflect adaptations to all users more broadly? The answer could have implications for the extent of inventorship accorded to individual researchers. These questions, and this very uncertainty, underscore the difficulty in applying traditional conception standards to joint human-AI inventions.

Finally, AI complicates nonobviousness under 35 U.S.C. § 103 (2018). If an AI arrives at a solution, but the AI's reasoning process is opaque or unavailable, can one *prove* a solution is non-obvious to a PHOSITA? A further question concerns the meaning of obviousness. If an AI can generate an innovation, does that imply the innovation is obvious? This line of reasoning follows from the argument that all AI outputs are merely recombinations of training data and prior inputs, and therefore that the AI's outputs cannot be truly novel.[109]

The crux of the issue, similar to authorship, arises from applying doctrines predicated on human conception to human-AI interactions where roles become deeply entangled. Assessing the legal significance of contributions is profoundly challenging when inputs and adaptations recursively shape each other, making separation difficult or impossible. With joint inventorship, the challenge is further compounded: the traditional task of delineating contributions among multiple human inventors—itself often complex—must now navigate the added complexities of a recursively adaptive AI.

Thus, the unmappability that confounds the allocation of creative rights in authorship proves just as disruptive when the law must allocate rights in inventorship. We now turn from these challenges facing authorship and inventorship to parallel crises in tort law.

---

[109] *See, e.g.*, Bender et al., *supra* note 34, at 617. Although, if so, this standard arguably should also apply to human-generated innovations: arguably, then an innovation should be deemed obvious if an AI can generate it without specific human guidance, even if it appears non-obvious to human experts.

## V. TORT LAW

Autonomy in AI systems has long raised legal and ethical challenges.[110] Two axes of control underpin current tort law as it relates to AI: (i) the degree of *user control over the AI's output*, and (ii) the *manufacturer's or developer's foreseeability of the AI's uses and potential harms*. When at least one axis is stable—that is, when either user control is high or manufacturer foreseeability is clear—the law can generally find a coherent path to allocate responsibility. As this Part will show, fluid agency in AI, however, destabilizes both at once: it weakens the causal link between particular actions and human decisions, and enables emergent actions rendering manufacturer foreseeability tenuous. These dynamics strain traditional liability frameworks predicated on tracing causation to assign responsibility.

To understand this collapse, we first ground the analysis in two baseline cases at the opposite ends of the agency spectrum. With traditional generative AI, liability frameworks largely adhere to a user-centric model.[111] Because the user maintains substantial control over outputs through iterative prompting and curation, legal responsibility typically falls on the human operator. For example, if a user employs ChatGPT to draft a legally binding contract that subsequently contains errors, courts will likely hold the user—not the AI or its developer—liable. The AI, in this context, is analogous to a sophisticated tool, like a word processor or a spreadsheet program, where the user directs the functionality and bears responsibility for the final product.[112]

At the opposite end of the spectrum lies fully autonomous AI systems, designed for independent decision-making without direct human oversight. Here, liability inquiries shift entirely to the manufacturer's foreseeability. The principle employed is that if an AI is designed for a specific, narrow purpose (e.g., a medical diagnostic tool), the manufacturer has greater foreseeability and thus a clearer duty to anticipate and mitigate risks. Yet, two distinct failure modes exist. First, it can lead to what Andreas Matthias terms the "responsibility gap": a scenario where an AI's actions, through learning or adaptation, extend beyond the foreseeable scope of its design, making it difficult to hold the manufacturer responsible, while the user's lack of control absolves them of direct liability.[113] Second, even when a general class of harm is foreseeable, the legal system's imperative to assign blame can lead to misallocation—human actors may be unfairly held accountable for AI decisions over which they had limited practical control, a phenomenon Madeleine Elish terms the "moral crumple zone."[114] This dynamic is illustrated in *People v. Riad*, where the driver of a Tesla using Autopilot pleaded no contest to vehicular manslaughter after a fatal crash, an outcome that underscores how legal culpability attaches to the human operator even when an advanced driver-assistance system is actively engaged.[115]

---

[110] *See* Peter M. Asaro, *A Body to Kick, but Still No Soul to Damn: Legal Perspectives on Robotics*, *in* ROBOT ETHICS: THE ETHICAL AND SOCIAL IMPLICATIONS OF ROBOTICS 169, 171 (Patrick Lin, Keith Abney & George A. Bekey eds., MIT Press 2012).

[111] This approach is consistent with principles of products liability, which generally limit a manufacturer's responsibility for harms arising from *not* reasonably foreseeable user misuse or post-sale modification. *See* RESTATEMENT (THIRD) OF TORTS: PRODUCTS LIABILITY § 2(b)–(c), cmt. m (discussing foreseeable risks) & cmt. p (discussing foreseeable misuse and modification) (AM. L. INST. 1998).

[112] *Cf.* BMG Music v. Gonzalez, 430 F.3d 888 (7th Cir. 2005) (holding a user liable for copyright infringement resulting from their use of file-sharing software, a tool similarly under their direct control).

These poles appear administrable only because each keeps at least one axis—control or foreseeability—stable. Fluid agency destabilizes both at once, weakening the direct causal link between a specific action and a human decision and rendering foreseeability fluid, making liability determinations profoundly challenging.[116]

Fluid agency breaks down these traditional anchors of liability through three distinct mechanisms. First, its epistemic opacity masks task complexity and creates the illusion of user control.[117] For instance, a non-expert relying on an AI code generator might be unaware of security flaws embedded within some generated code. If that code is then deployed and exploited, the user could face disproportionate liability for vulnerabilities they could not reasonably have detected or prevented.[118] This problem is compounded by regulatory frameworks that, in seeking to enhance transparency, inadvertently reinforce this illusion. The EU's General Data Protection Regulation (GDPR), for example, provides certain explanation-adjacent rights.[119] Even if users cannot practically access and understand these model explanations, the mere existence of explanation-adjacent rights risks may create a *de facto* presumption of user control, potentially exposing users to liability for harms they could neither foresee nor prevent.

---

[113] *See* Matthias, *supra* note 20, at 177; *see also* Filippo Santoni de Sio & Giulio Mecacci, *Four Responsibility Gaps with Artificial Intelligence: Why They Matter and How to Address Them*, 34 PHIL. & TECH. 1057, 1060 (2021).

[114] *See* Elish, *supra* note 51, at 44.

[115] *See* ASSOCIATED PRESS, *As a criminal case against a Tesla driver wraps up, legal and ethical questions on Autopilot endure* (Aug. 15, 2023), https://apnews.com/article/tesla-autopilot-los-angeles-d02769ba359cf6381dc1176c3f5a72a5 [https://perma.cc/9LCM-54FE] (reporting on the no-contest plea and sentence in People v. Riad, No. TA155613 (Cal. Super. Ct., L.A. Cnty.)).

[116] Recognizing this principle, the EU AI Act imposes detailed, *ex ante* obligations on providers of "high-risk" AI systems, with complementary duties for deployers, requiring extensive risk assessment, data governance, and human oversight to mitigate the potential for unforeseen harms. *See* Regulation (EU) 2024/1689, *supra* note 71, arts. 9–10, 14–15, 16 (obligations of providers), 26 (obligations of deployers).

[117] *See, e.g.*, Frank Pasquale, *The Black Box Society: The Secret Algorithms That Control Money and Information*, HARVARD UNIV. PRESS (2015) (analyzing the societal and legal

Second, recursive human–machine co-production erases attribution. The interplay between users and AI systems makes it exceedingly difficult, if not impossible, to disentangle contributions. Consequently, it can become impossible to definitively state whether a particular output stems from direct user instruction, the AI's autonomous decision-making, or an inseparable fusion of both, collapsing the causal chain needed for legal attribution.[120]

---

challenges posed by opaque algorithmic systems); *see also* Asaro, *supra* note 110, at 171 (discussing tools that "mask their own complexity").

[118] This liability may be shared with the manufacturer or developer. For instance, the EU's recast Product Liability Directive establishes a strict liability regime for defective products, expressly clarifying that software—including AI systems—is a "product" and treating AI providers as "manufacturers." *See* Directive (EU) 2024/2853 of the European Parliament and of the Council of 23 October 2024 on liability for defective products, O.J. (L 2024/2853) 18.11.2024, art. 4(1) (establishing that a producer can be held liable for damage caused by a defect in its product, including software, without proof of negligence). The AI Act, in contrast, imposes *ex ante* regulatory obligations rather than its own civil liability rules. *See* Regulation (EU) 2024/1689, *supra* note 71.

[119] *See* Regulation (EU) 2016/679, arts. 13–15, 22 & recital 71, 2016 O.J. (L 119) 1 (providing for rights to "meaningful information about the logic involved" in automated decision-making and safeguards against decisions based solely on such processing).

[120] While the GDPR's oft-invoked "right to an explanation" is not a freestanding right and its scope is highly contested, the existence of disclosure requirements under Articles 13-15 and safeguards under Article 22 creates a veneer of user empowerment that may not reflect operational reality. *See supra* note 119; *see also* Sandra Wachter, Brent Mittelstadt & Luciano Floridi, *Why a Right to Explanation of Automated Decision-Making Does Not Exist in the General Data Protection Regulation*, 7 INT'L DATA PRIVACY L. 76 (2017).

Third, the AI's capacity for "function creep"[121] creates unstable foreseeability for the manufacturer. For example, an AI initially designed for legal contract drafting might, through user interaction and adaptation, evolve to perform tasks far beyond its original intended scope, such as financial forecasting. Such fluidity of purpose can make it difficult to apply traditional liability frameworks that rely on a clear distinction between intended and unintended uses as the very concept of a fixed "intended use" becomes meaningless when the system's capabilities are dynamic.

This breakdown of individual liability escalates at the enterprise level, where the doctrine of vicarious liability becomes an awkward fit. For instance, organizational deployment of AI destabilizes *respondeat superior*, which holds employers liable for harms caused by employees acting within the scope of employment. The doctrine presupposes an employer-employee relationship and hinges on an employer's right to control an employee's actions. As an AI is not a person, it can neither be a legal "employee" nor a more general "agent" rendering the doctrine inapplicable.[122] At the same time, U.S. anti-discrimination law squarely places liability on employers for the use of algorithmic tools that cause disparate impact, even if a vendor built the tool.[123] Consider an AI hiring agent that autonomously screens job applicants. If this agent develops discriminatory patterns through recursive adaptation, courts face an attribution paradox. The AI's behavior may reflect neither explicit corporate policy nor any individual employee's intent. Because the AI's actions are autonomous and potentially unforeseeable, the employer may lack the requisite control to be held liable under *respondeat superior*. Yet, it directly causes harm.

This is a novel systemic *enterprise responsibility gap*: a situation distinct from Matthias's general gap and from the moral crumple zone. The organization benefits economically from the AI's actions but may evade liability for the resulting harms because the unmappability of causation—whether the bias originated in training data, deployment

---

[121] *See* Arif Kornweitz, *A New AI Lexicon: Function Creep*, AI NOW INST. (Aug. 4, 2021) (defining function creep as "the expansion of the functionality of an algorithmic system, a divergence of its initial purpose and actual use," and arguing that this is not an exception but the "modus operandi" of such systems).

[122] The RESTATEMENT (THIRD) OF AGENCY § 1.01 (AM. L. INST. 2006) defines agency as a fiduciary relationship between two "persons"—a principal and an agent.

[123] *See* U.S. EQUAL EMP. OPPORTUNITY COMM'N, SELECT ISSUES: ASSESSING ADVERSE IMPACT IN SOFTWARE, ALGORITHMS, AND AI USED IN EMPLOYMENT SELECTION PROCEDURES UNDER TITLE VII OF THE CIVIL RIGHT ACT OF 1964 (2023) (clarifying that employers may be liable for the use of vendor-provided tools); U.S. DEP'T OF JUST. & U.S. EQUAL EMP. OPPORTUNITY COMM'N , ALGORITHMS, ARTIFICIAL INTELLIGENCE, AND DISABILITY DISCRIMINATION IN HIRING (2022) at 12.

parameters, or autonomous adaptation—leaves no clear doctrinal path to hold the enterprise accountable.[124]

In sum, fluid agency creates at least three distinct dead zones: (1) the *responsibility gap*, where no party fits the doctrinal seat for liability *ex post*; (2) the *moral crumple zone*, where liability is not absent but is misallocated, defaulting to the nearest human operator despite their limited control; and (3) novel to fluid agency, the *enterprise responsibility gap*, where an organization benefits from an AI's autonomous actions but evades liability because the fundamental unmappability leaves no doctrinal hook.

These failures share a common pathology: the breakdown of attribution-based frameworks when confronted with irreducibly entangled human-AI processes. A workable framework must satisfy several key design imperatives: it must align liability with the party best positioned for *ex ante* risk control, guarantee compensability for harms in high-impact domains, create process-based safe harbors for diligent actors, and ensure that responsibility is allocated in a way that is both proportional and supported by legible evidence.

Any workable solution must therefore abandon intractable causal inquiries in favor of a new logic: allocating rights and responsibilities based on outcomes and *ex ante* control, rather than source. This is the principle of functional equivalence that we delineate in the next Part.

# VI. THE PRINCIPLE OF FUNCTIONAL EQUIVALENCE

This Article proposes a paradigm shift: treating human and AI contributions as *functionally equivalent*, not because of moral or economic parity, but as a pragmatic response to the *fundamental unmappability* of contributions and control. By "functional equivalence," we mean that legal frameworks should focus on the *qualities and outcomes* of human-AI interactions rather than pursuing an intractable inquiry into their origins. This shift is necessary to circumvent three persistent problems: (1) consistently determining *when* contributions can even be parsed; (2) the absence of fair or workable standards for *partial attribution*; and (3) *inequities* from treating similar collaborations differently based on happenstance traceability. Before detailing how this principle applies across legal domains, we summarize why more familiar alternatives are inadequate.

---

[124] Emerging regulation is beginning to address this accountability challenge. In July 2025, the California Privacy Protection Agency board voted to finalize regulations governing Automated Decisionmaking Technology (ADMT), and on September 23, 2025, the Office of Administrative Law approved them; the rules take effect January 1, 2026 (with ADMT requirements beginning January 1, 2027). By creating affirmative duties of transparency and consumer control (*e.g.*, pre-use notices, access, and opt-out rights) for businesses deploying these systems, this rulemaking aims to mitigate the enterprise responsibility gap by shifting the focus from intractable causal attribution to process-based accountability. *See* CAL. CIV. CODE § 1798.185(a)(16) (West 2024); *see also* CAL. PRIV. PROT. AGENCY, *California Finalizes Regulations to Strengthen Consumers' Privacy* (Sep. 23, 2025); Caitlin Andrews, *CPPA Board Finalizes Long-Awaited ADMT, Cyber Audit, Risk Assessment Rules*, INT'L ASS'N OF PRIV. PROS. (July 25, 2025).

Why familiar alternatives fail under unmappability

- **Hybrid authorship/joint inventorship.** Requires partitioning contributions; collapses when contributions are inseparable. Adding subjective tests like "intent" or "meaningful control" (for authorship), or "collaborative connection" (for inventorship), reimports unworkable line-drawing.

- **Proportional royalties/weights.** Needs a reliable metric for relative contribution; with *fluid agency*, weights are unobservable and contestable, leading to litigation over proxies (e.g., edit counts, logs).

- **Provenance fixes (logs, watermarking).** Track inputs and outputs, not internal adaptation; cannot evidence conceptual origin in recursive loops.

- ***Sui generis* output right.** Avoids the origin problem but breaks the essential link between rights and responsibility (e.g., who bears the duty to warn or recall?) and risks flooding markets with low-cost, legally protected output.

- **Fault-based tort allocation.** Ineffective in *moral crumple zones*, where fault is misallocated to the nearest human even when their control was nominal. Strict or no-fault enterprise allocations better tie responsibility to the party best positioned to manage risk ex ante.

For copyright, functional equivalence would be implemented through a rebuttable presumption of human authorship. When reasonable provenance practices are followed but fail to yield a separable attribution, the law would presume that the final work's protectable authorship vests in the human orchestrator or commissioning entity. This presumption would primarily concern registrability and evidentiary weight in enforcement, not the subsistence of the right itself, thereby respecting international norms such as the Berne Convention's prohibition on formalities. To benefit from this presumption, an applicant would need to (i) identify the human initiator and orchestrator, (ii) maintain and be prepared to produce process records (e.g., prompt and edit logs, model and version disclosures), and (iii) certify final human review and approval of the work. The presumption could be rebutted by clear evidence of purely autonomous generation without meaningful human orchestration.[125]

In patent law, functional equivalence would require a normative reform to address inventions where the pivotal insight is AI-generated.[126] Under this principle, patentability (novelty, non-obviousness, and utility) would be assessed based on the invention's objective merits, but inventorship would be granted to the human(s) who (i) orchestrated the AI-assisted discovery process, (ii) verified and validated the result, and (iii) effected reduction to practice. This approach resolves the *DABUS* impasse by focusing on the human role in delivering the invention to the public, subsuming the AI's conceptual contribution within the human-directed workflow. To satisfy disclosure requirements, the human inventor would need to demonstrate possession of the claimed subject matter by articulating its operative parameters and replication procedures—not merely by pointing to a successful "black box" output—and would have a duty to disclose the AI's material involvement to the patent office. This logic, which rewards human orchestration, finds a parallel in the reasoning of cases like *Shenzhen Tencent*, where creative rights were allocated based on the human team's role in guiding the AI process.[127]

Liability models, under functional equivalence, would shift from intractable causal inquiries to frameworks that align responsibility with *ex ante* control (design and deployment choices). In high-impact or safety-critical domains, this may point toward activity-based strict liability for developers and expanded enterprise liability for the organizations that deploy AI systems.[128] This responsibility, however, could be mitigated by a process-based safe harbor, where liability is reduced for entities that can demonstrate adherence to best practices in risk assessment, pre-deployment testing, post-deployment monitoring, and diligent oversight. For widespread, lower-level harms, sector-specific no-fault compensation schemes—akin to the National Vaccine Injury Compensation Program—could ensure redress without requiring a plaintiff to parse an unmappable causal chain.[129] Alternatively, a more doctrinally conservative path is to develop new evidentiary tools that make traditional fault-based inquiries tractable again. Such an approach would require a structured method for analyzing an AI's operational characteristics—its goals, predictive capacities, and safety features—as a proxy for human fault.[130]

---

[125] This outcome resembles statutory solutions designed for earlier forms of machine creation, such as the UK's rule for computer-generated works, which vests authorship in the person who makes the necessary arrangements for the work's creation. *See* Copyright, Designs and Patents Act 1988, c. 48, §§ 9(3), 178 (UK). The principle of functional equivalence helps explain why such outcome-focused approaches are workable: they implicitly address the challenge of attribution, a logic that becomes even more critical in the face of the unmappability caused by fluid agency.

[126] Current U.S. law requires an inventor to be a natural person, a holding that necessitates statutory reform to implement an orchestration-based standard. *See supra* notes 99 and 103.

[127] *See* Shenzhen Tencent Comput. Sys. Co. v. Shanghai Yingxun Tech. Co., China Justice Observer (Shenzhen Nanshan Dist. People's Ct. Dec. 24, 2019) (China).

[128] Such proposals build on the long tradition of adapting tort law to place liability on the party best positioned to manage risk. *See, e.g.*, Escola v. Coca Cola Bottling Co. of Fresno, 24 Cal. 2d 453, 462–64 (1944) (Traynor, J., concurring) (providing the intellectual

Critics may legitimately argue that functional equivalence risks diminishing the perceived value of human creativity or that it is a step toward allocating rights to machines, which do not need incentives to create.[131] Our proposal, however, does not grant rights to machines; it allocates them to the human orchestrator as a pragmatic necessity. While acknowledging these concerns, we contend that legal frameworks must prioritize administrability. The legal system has historically evolved to address technological shifts: corporate personhood enabled firms to act as legal entities without equating them to

---

foundation for strict products liability); Ira S. Bushey & Sons, Inc. v. United States, 398 F.2d 167, 171 (2d Cir. 1968) (expanding vicarious liability based on an enterprise's foreseeable risks); GUIDO CALABRESI, THE COSTS OF ACCIDENTS: A LEGAL AND ECONOMIC ANALYSIS 135 (Yale University Press 1970) (arguing that liability should be allocated to the cheapest cost avoider).

[129] *See* National Childhood Vaccine Injury Act, 42 U.S.C. §§ 300aa-10 to -3 (201820183 (West)). The law has repeatedly created such pragmatic frameworks when traditional models fail. *See, e.g.*, Directive 96/9/EC, *supra* note 72, art. 7. The common thread is the replacement of intractable attribution with administrable, outcome-focused mechanisms.

[130] For a full development of such a framework, *see* Anirban Mukherjee & Hannah H. Chang, *Operational Agency: A Framework for Tracing Intent and Liability in Multi-Agent Artificial Intelligence Systems* (Oct. 12, 2025), https://ssrn.com/abstract=5344615. [http://dx.doi.org/10.2139/ssrn.5344615].

[131] *See, e.g.*, Joanna J. Bryson, *Robots Should Be Slaves, in* CLOSE ENGAGEMENTS WITH ARTIFICIAL COMPANIONS 63 (Yorick Wilks ed., 2010) (arguing that granting rights to robots could diminish human status); Carys J. Craig & Ian R. Kerr, *The Death of the AI Author*, 52 OTTAWA L. REV. 33, 43 (2021) (quoting Samuelson, *supra* note 55, at 1199 ("[I]t simply does not make any sense to allocate intellectual property rights to machines because they do not need to be given incentives to generate output.")).

human moral agents; copyright expanded to protect photographs and software without demanding proof of "humanity" in each pixel or line of code. The system must now confront creative processes where human and AI contributions are irreducibly entangled. In such cases, distinctions based on origin are not merely difficult, but impractical to apply fairly. Our focus on outcomes, embodied in functional equivalence, stems not from a philosophical preference but from the necessity of maintaining a workable legal framework.

With the principle of functional equivalence established and defended, we now turn to concrete pathways for its implementation and the broader questions that remain.

# VII. CONCLUSION: FROM ORIGINS TO OUTCOMES

Modern AI systems exercise fluid agency. Their stochastic, dynamic, and adaptive workflows entangle human and machine contributions so tightly that the provenance of creative works and harmful acts becomes fundamentally unmappable. Foundational legal doctrines that depend on tracing origins fracture—not because they are normatively misguided, but because they presuppose a clear chain of attribution that no longer exists.

This Article has argued for a pragmatic default to resolve this doctrinal collapse: functional equivalence, which treats human and AI contributions as equivalent for legal allocation. The claim is not one of moral or ontological parity between persons and machines. Instead, it is an administrable principle that preserves incentives and accountability by shifting the inquiry from an unworkable search for origins to a functional assessment of who orchestrates the process, validates the outcome, and is best positioned *ex ante* to manage risk.

Implementing this principle would necessitate legislative and judicial recalibration.[132] Registration procedures could adopt a rebuttable presumption of human authorship, absent clear evidence of purely autonomous generation, balancing incentives for human creativity with the reality of blended contributions. Streamlined registration processes—perhaps similar to the U.S. Copyright Office's group registration options—[133] could acknowledge the collaborative nature of works without demanding precise attribution. Legislative action or judicial reinterpretation of related case law could recognize inventorship based on human supervision and reduction to practice, even when the pivotal conception originated with AI. This approach would depart from the European Patent Convention's strict adherence to human inventorship[134] and support the contention that AI capabilities exceeding human expertise warrant rethinking traditional human-centered standards like the PHOSITA benchmark. Policymakers might adopt a strict liability model for developers of "high-risk" AI systems (as designated in regimes such as the EU AI Act),[135] while retaining user responsibility for foreseeable misuse under established negligence principles. This hybrid approach echoes prior calls for structured solutions that move beyond a simple application of traditional tort doctrines, which are often ill-suited to AI's unpredictability.[136]

---

[132] The stakes of such recalibration are high, as current litigation involves not only staggering statutory damages but also demands for the outright destruction of AI models. *See* Pamela Samuelson, *How to Think about Remedies in the Generative AI Copyright Cases*, 67 COMMUN. ACM 27, 27 (2024).

[133] *See e.g., Circular 42: Copyright Registration of Photographs*, U.S. COPYRIGHT OFFICE: CIRCULARS (Mar. 2021), https://www.copyright.gov/circs/circ42.pdf [https://perma.cc/C8LS-CAK3].

This analysis has limitations. Our focus on U.S. law leaves open crucial questions of comparative jurisprudence. Most notably, how will civil law systems reconcile fluid agency with statutory obligations, such as the EU AI Act's mandates for human oversight (Art. 14) and transparency (Art. 13)?[137] While U.S. law grapples *post hoc* with the attribution challenges arising from unmappability, the EU AI Act's emphasis on *ex ante* risk assessment and conformity requirements[138] may preemptively constrain fluid agency,

---

[134] *See* European Patent Convention, Convention on the Grant of European Patents, Oct. 5, 1977, 1065 U.N.T.S. 199 r. 19(1).

[135] *See* Council Regulation 2024/1689, *supra* note 71.

[136] *See* Omri Rachum-Twaig, *Whose Robot Is It Anyway?: Liability for Artificial-Intelligence-Based Robots*, 2020 U. ILL. L. REV. 1141, 1143–44. While Rachum-Twaig proposes a different mechanism—a 'presumed negligence' framework triggered by failing specific 'safe harbor' duties (e.g., monitoring, patching)—the underlying goal of establishing clearer responsibility benchmarks for developers and users aligns with the functional equivalence approach advocated here.

[137] *See* Council Regulation 2024/1689, *supra* note 71.

[138] *See* Council Regulation 2024/1689, *supra* note 71 at art. 43 (mandating a significant *ex ante* verification regime: many high-risk AI systems must undergo conformity assessments—some by third parties—*before* being placed on the market or put into service). This pre-market scrutiny, focusing on transparency, safety, and fundamental rights, represents a fundamentally different regulatory philosophy compared to legal systems relying primarily on *post hoc* liability determination after harm has occurred. For the legislative intent behind this *ex ante* approach, *see Proposal for a Regulation Laying Down Harmonised Rules on Artificial Intelligence (Artificial Intelligence Act)*, at 9–11, COM

trading some adaptive potential for clearer accountability. Whether such a regulatory bargain proves sustainable—and whether it offers lessons for common law systems—remains an open question.

Our arguments rest on capabilities that are still crystallizing, not on technologies whose properties are already well understood. Two complementary lines of investigation are therefore crucial. First is quantitative research: large-scale logging of human–AI interaction data can be paired with econometric and information-theoretic tools, such as causal inference models and Shapley value decompositions, to estimate how much of a given artifact is best explained by human prompts versus AI's autonomous processes. The goal is not to achieve perfect provenance, but to supply courts with statistical evidence analogous to what they already accept when apportioning antitrust damages or detecting employment discrimination.[139] Second is qualitative inquiry: ethnographic studies could explore how engineers, legal professionals, and artists perceive agency within human-AI co-productions, revealing whether folk norms of attribution converge with or resist emergent legal doctrine. Taken together, these empirical programs would translate "unmappability" from an abstract idea into both measurable (if fuzzy) quantities and lived experiences. Courts would gain a familiar evidentiary scaffold on which to build doctrine, grounding the pragmatic case for functional equivalence.

---

(2021) 206 final (Apr. 21, 2021). For an analysis of how the act shifts compliance burdens to earlier stages of the AI lifecycle, *see* Michael Veale, Kira Matus & Robert Gorwa, *AI and Global Governance: Modalities, Rationales, Tensions*, 19 ANN. REV. L. & SOC. SCI. 255 (2023).

[139] For examples of technical attribution methods, *see* Scott M. Lundberg & Su-In Lee, *A Unified Approach to Interpreting Model Predictions*, *in* 30 ADVANCES IN NEURAL INFO. PROCESSING SYS. 4765–74 (Sanmi Koyejo et. al. eds., 2017) (introducing SHAP values); JUDEA PEARL, CAUSALITY: MODELS, REASONING, AND INFERENCE (2d ed. 2009). Courts have a long and established history of relying on such complex, probabilistic evidence. *See, e.g.*, Wal-Mart Stores, Inc. v. Dukes, 564 U.S. 338, 356–57 (2011) (discussing statistical and anecdotal evidence in class certification); Int'l Bhd. of Teamsters v. United States, 431 U.S. 324, 339 (1977) (recognizing the probative value of statistical proof in "pattern or practice" cases); Daniel L. Rubinfeld, *Econometrics in the Courtroom*, 85 COLUM. L. REV. 1048 (1985). Such models might, for example, weight human inputs (e.g., prompts, edits) against AI-generated variations to determine when disentanglement becomes impractical—a challenge courts already navigate in fuzzy determinations like "substantial similarity" or "non-obviousness." *See* Arnstein v. Porter, 154 F.2d 464, 468 (2d Cir. 1946); *see also* KSR Int'l Co. v. Teleflex Inc., 550 U.S. 398, 415–22 (2007) (emphasizing a "flexible" approach to the obviousness inquiry).

These empirical needs point to a deeper, normative quandary that transcends doctrinal mechanics: a challenge to intellectual property law's cultural mission. When AI systems trained on existing works replicate and recombine dominant aesthetic and narrative patterns, they risk creating a feedback loop that homogenizes culture. This process can not only marginalize nonconforming styles and viewpoints but also erode the expressive diversity that copyright law is meant to foster.[140] Similarly, the patent system justifies its grant of exclusive rights on the premise that it incentivizes costly, high-risk, and nonobvious invention that would not otherwise occur.[141] The principle of functional equivalence may alter this calculus. If AI's computational power makes certain types of innovation cheap and predictable, its algorithmic path dependencies may steer invention toward those pathways rather than disruptive breakthroughs. This could lead to patent portfolios that are deep but narrow, rewarding rapid, derivative works in specific domains over foundational research broadly.

In sum, an uncritical embrace of the principle of functional equivalence risks inadvertently calcifying systemic inequities behind a veneer of technological progress, undermining the very innovation and diversity our legal frameworks were built to protect. However, even as we remain mindful of this risk, we argue that our legal systems must evolve beyond purely anthropocentric paradigms to embrace functional equivalence as a practical necessity. Maintaining the status quo would create a legal landscape that either stifles technological progress by adhering to unworkable standards or fails to adequately protect the human authors and innovators whom the legal system seeks to serve. By shifting focus from unmappable origins to tangible outcomes, the principle of functional equivalence offers a more secure foundation for allocating rights coherently and assessing liability fairly.

---

[140] The goal of copyright is not merely to incentivize the production of more works, but to "promote the Progress of Science and useful Arts," U.S. CONST. art. I, § 8, cl. 8, a purpose long understood to include the enrichment of public discourse and culture. *See, e.g.*, Jack M. Balkin, *Digital Speech and Democratic Culture: A Theory of Freedom of Expression for the Information Society*, 79 N.Y.U. L. REV. 1, 35–38 (2004); Julie E. Cohen, *Copyright, Commodification, and Culture: Locating the Public Domain*, *in* THE FUTURE OF THE PUBLIC DOMAIN: IDENTIFYING THE COMMONS IN INFORMATION LAW 121, 136–40 (Lucie Guibault & P. Brent Hugenholtz eds., 2006).

[141] *See, e.g.*, Graham v. John Deere Co. of Kan. City, 383 U.S. 1, 6 (1966) (linking the patent monopoly to the constitutional goal of promoting invention).